\documentclass[twocolumn]{aastex62}

\usepackage{graphicx}
\usepackage{longtable}
\usepackage{amssymb, amsfonts}
\usepackage{amsmath}
\usepackage{booktabs}
\usepackage{tablefootnote}
\usepackage{multirow}

\begin{document}
\def\hh{\, h^{-1}}
\newcommand{\ie}{$i.e.$,}
\newcommand{\wth}{$w(\theta)$} 
\newcommand{\mpc}{Mpc}
\newcommand{\xir}{$\xi(r)$}
\newcommand{\Lya}{Ly$\alpha$}
\newcommand{\Lyb}{Lyman~$\beta$}
\newcommand{\Hb}{H$\beta$}
\newcommand{\msun}{M$_{\odot}$}
\newcommand{\hmsun}{$h^{-1}$M$_{\odot}$}
\newcommand{\sfr}{M$_{\odot}$ \text{yr}$^{-1}$}
\newcommand{\dnsty}{$h^{-3}$Mpc$^3$} 
\newcommand{\za}{$z_{\rm abs}$}
\newcommand{\ze}{$z_{\rm em}$}
\newcommand{\cmtwo}{cm$^{-2}$}
\newcommand{\nhi}{$N$(H$^0$)}
\newcommand{\degpoint}{\mbox{$^\circ\mskip-7.0mu.\,$}}
\newcommand{\halpha}{\mbox{H$\alpha$}}
\newcommand{\hbeta}{\mbox{H$\beta$}}
\newcommand{\hgamma}{\mbox{H$\gamma$}}
\newcommand{\kms}{\,km~s$^{-1}$}      
\newcommand{\minpoint}{\mbox{$'\mskip-4.7mu.\mskip0.8mu$}}
\newcommand{\mv}{\mbox{$m_{_V}$}}
\newcommand{\Mv}{\mbox{$M_{_V}$}}
\newcommand{\peryr}{\mbox{$\>\rm yr^{-1}$}}
\newcommand{\secpoint}{\mbox{$''\mskip-7.6mu.\,$}}
\newcommand{\sqdeg}{\mbox{${\rm deg}^2$}}
\newcommand{\squig}{\sim\!\!}
\newcommand{\subsun}{\mbox{$_{\twelvesy\odot}$}}
\newcommand{\et}{{\it et al.}~}
\newcommand{\er}[2]{$_{-#1}^{+#2}$}
\def\h50{\, h_{50}^{-1}}
\def\hbl{km~s$^{-1}$~Mpc$^{-1}$}
\def\ltsima{$\; \buildrel < \over \sim \;$}
\def\simlt{\lower.5ex\hbox{\ltsima}}
\def\gtsima{$\; \buildrel > \over \sim \;$}
\def\simgt{\lower.5ex\hbox{\gtsima}} 
\def\arcs{$''~$}
\def\arcm{$'~$}
\newcommand{\wu}{$U$}
\newcommand{\wb}{$B_{435}$}
\newcommand{\wv}{$V_{606}$}
\newcommand{\wi}{$i_{775}$}
\newcommand{\wz}{$z_{850}$}
\newcommand{\hmpc}{$h^{-1}$Mpc}
\newcommand{\lm}{$L$--$M$}
\newcommand{\ws}{$\mathcal{S}$}
\newcommand{\wm}{$\mathcal{M}$}
\newcommand{\sm}{$\mathcal{S}$-$\mathcal{M}$}
\newcommand{\medianLM}{$\tilde{\mathcal{L}}(M)$}
\newcommand{\mf}{$\phi_\mathcal{M}$}
\newcommand{\ebv}{$E$($B-V$)}
\newcommand{\fesc}{$f_{\rm esc}$}
\makeatletter
\newcommand*{\rom}[1]{\expandafter\@slowromancap\romannumeral #1@}
\makeatother

\defcitealias{shi19b}{S19}
\defcitealias{xue17}{X17}

\title{The Role of Dust, UV Luminosity and Large-scale Environment on the Escape of Ly$\alpha$ Photons:\\ A Case Study of a Protocluster field at $z=3.1$}

\author{Yun Huang}
\affiliation{Department of Physics and Astronomy, Purdue University, 525 Northwestern Avenue, West Lafayette, IN 47907}
\author{Kyoung-Soo Lee}
\affiliation{Department of Physics and Astronomy, Purdue University, 525 Northwestern Avenue, West Lafayette, IN 47907}
\author{Ke  Shi}
\affiliation{Department of Astronomy, Xiamen University, Xiamen, Fujian 361005, People’s Republic of China}
\author{Nicola Malavasi}
\affiliation{Universit\'{e} Paris-Saclay, CNRS, Institut d'Astrophysique Spatiale, 91405, Orsay, France}
\author{Rui Xue}
\affiliation{National Radio Astronomy Observatory, 520 Edgemont Road, Charlottesville, VA 22903, USA}
\author{Arjun Dey}
\affiliation{NSF's National Optical-Infrared Astronomy Research Laboratory, 950 N. Cherry Ave., Tucson, AZ 85719, USA}

\begin{abstract}
We present a detailed characterization of the Ly$\alpha$ properties for 93 Ly$\alpha$ emitters (LAEs) at $z\sim3.1$ selected from the D1 field of the Canada-France-Hawaii-Telescope Legacy Survey, including 24 members of a massive protocluster. 
The median-stacked Ly$\alpha$ image
shows an extended Ly$\alpha$ halo (LAH) surrounding the galaxy with the exponential scale length $4.9\pm0.7$~kpc, which accounts for roughly half of the total line flux. Accounting for the LAH contribution, the total Ly$\alpha$ escape fraction, $f_{\rm esc}$, is $40\pm26$\%. 
Combining the dataset with existing measurements, we find a dependence of \fesc\ on the galaxy's UV slope ($\beta$) and UV luminosity ($L_{\rm UV}$).
The simultaneous use of both parameters allows prediction of \fesc\ within 0.18~dex, a substantial improvement over 0.23~dex when only $\beta$ is used. 
The correlation between \fesc\ and E$(B-V)$ suggests that Ly$\alpha$ photons undergo interstellar dust attenuation in a similar manner to continuum photons. Yet, Ly$\alpha$ transmission is typically higher than that expected for continuum photons at similar wavelength by a factor, which depends on UV luminosity, up to 2 in the samples we studied. 
These results hint at complex geometries and physical conditions of the interstellar medium, which affect the Ly$\alpha$ transmission or production. Alternatively, the dust law may change with luminosity leading to over-or under-estimation of \fesc. 
Finally, we report that protocluster LAEs tend to be bluer and more UV-luminous than their field cousins, resulting in systematically higher \fesc\ values. We speculate that it may be due to the widespread formation of young low-mass galaxies in dense gas-rich environments. 
\end{abstract}


\section{Introduction} \label{sec:intro}

Ly$\alpha$ emission plays a  central and multifaceted role in our understanding the formation and evolution of galaxies in the distant universe. Existing studies show that galaxies identified via their strong Ly$\alpha$ emission \citep[Ly$\alpha$ emitters or LAEs, hereafter: see review by][]{ouchi20} tend to be blue, UV-faint, and low-mass galaxies \citep{gawiser06,finkelstein07,guaita11}, representing a  population of galaxies that not only  drives the cosmic star formation rate density at  $z\gtrsim 2$ \citep{reddy09} but also is the  primary building blocks of typical present-day galaxies such as our own Milky Way \citep[e.g.,][]{gawiser07}.
 
As a resonant line, the properties of Ly$\alpha$ emission, such as the escape fraction and the spectral shape, offer vital clues to  the spatial and kinematic structures of the neutral gas and dust in the interstellar and circumgalactic media \citep[ISM and CGM, respectively: e.g.,][]{verhamme12,dijkstra12,rivera15}. Studies  found that the escape of Ly$\alpha$ photons strongly and positively correlates with the escape of Lyman continuum (LyC) photons in low-redshift analogs \citep[e.g.,][]{verhamme17,gazagnes20} and in high-redshift galaxies \citep[e.g.,][]{steidel18}, lending credence to the possibility that understanding  Ly$\alpha$ emission in galaxies may be key to constraining the reionization of the universe. 

Deep narrow-band imaging surveys have recently demonstrated the utility of Ly$\alpha$ emission as tracers of the large-scale structure in the distant universe. Massive protocluster sites are found to be significant overdensities of LAEs \citep{steidel00,lee14,badescu17,shi19b}. In several protoclusters,  filamentary structures traced by LAEs stretch out tens of Mpcs from a protocluster \citep[][]{matsuda05,dey16}, mirroring the expectation of  dark matter structures around a cluster-sized halo.
\citet{chiang17} noted that, as pre-virialized and highly overdense structures, protoclusters and the galaxies therein play a significant role in driving the cosmic star formation rate density of the universe at high redshift. To better quantify their importance, however, more detailed understanding of how galaxy formation proceeds in high-density environment is required. Although recent studies are beginning to address these questions, there is no clear consensus yet \citep[see, e.g., ][]{shi19a,shi20,shi21,lemaux20,malavasi21}.

Together with the current and upcoming deep wide-field optical imaging surveys employing both broad- and narrow-band filters \citep[e.g., the Subaru Strategic Program with the Hyper Suprime Cam, the Legacy Survey of Space and Time with the Vera C. Rubin telescope, SILVERRUSH, and ODIN:][also see Section~\ref{sec:environment}]{lsst,ouchi18,ssp}, these considerations make it likely that Ly$\alpha$ emitting galaxies will continue to play a crucial role in our understanding of the cosmos well into the future. 

One crucial diagnostic that is central to elucidating the  key questions mentioned above is Ly$\alpha$ escape fraction (\fesc, hereafter), which encapsulates the prominence of the Ly$\alpha$ relative to the stellar light. Thus far, the intrinsic faintness of most LAEs has limited the \fesc\ measurements only to those residing in a handful of deep surveys \citep[e.g.][]{blanc2011,song14,oyarzun17,vandels}, resulting in relatively small number statistics and a narrow dynamic range in galaxy properties. 

To further complicate matters, there is a growing consensus that the presence of an extended low surface brightness Ly$\alpha$ emission (often referred to as Lyman Alpha Halos or LAHs) is ubiquitous around star-forming galaxies at both low-  \citep[e.g.,][]{hayes13} and high redshift \citep[e.g.,][]{rauch08,steidel11,matsuda12,momose14,momose16,wisotzki,xue17,lecle17,kusakabe18}. The dominant process responsible for the production the LAH remains elusive as is  the relative importance of the LAH in different galaxy types \citep[see][]{momose16,xue17}. If the LAH phenomenon is produced by the photons originating from the H~{\sc ii} regions,  the true Ly$\alpha$ escape fraction may be considerably larger than currently known. 
 In this context, the existence of an LAH potentially broadens the scope and the importance of Ly$\alpha$ emission in the galaxy formation research.

In this paper, we study how the Ly$\alpha$ escape fraction of a galaxy changes with its physical properties and  large-scale environment. 
In Section~\ref{sec:obs}, we provide a brief description of our primary, ancillary datasets, and the LAE sample selection, and give the definition of the protocluster- and field  subsamples. 
In Section~\ref{sec:lah}, we present our LAH measurements and comparison to the literature. 
In Sections \ref{sec:esc} ad \ref{sec:dust}, we examine how \fesc\ varies with the galaxy's photometric properties  and present an empirical formula that can predict a galaxy's \fesc\ given other parameters; the implications of our results on the physical conditions of the ISM/CGM are also discussed. Section~\ref{sec:environment} explores how large-scale environment affects \fesc.
We summarize the results in Section~\ref{sec:con}.

Throughout this paper, we adopt the cosmology with $\Omega=0.27$, $\Omega_{\Lambda}=0.73$, $H_0=70$ km~s$^{-1}$~Mpc$^{-1}$ \citep{komatsu11}. Distance scales are given in comoving units unless noted otherwise. All magnitudes are given in the AB system \citep{oke83}.

\section{Observation and Sample Selection} \label{sec:obs}

Our LAE sample at $z=3.13$ is selected from one of the four  Canada-France-Hawaii-Telescope Legacy Survey deep fields  \citep[CFHTLS D1 field, hereafter:][]{gwyn12}. The details of the imaging dataset and the selection of our LAE sample at $z=3.13$ are given in \citet[hereafter S19]{shi19b}, and here we only provide a brief summary. 
The narrow-band observation is obtained using the Mosaic 3 Camera \citep{dey16b} at the Mayall 4m telescope using the KPNO \#k1014 filter ($o3$ filter, hereafter) with a central wavelength of 5024.9\AA\ and a full-width-at-half-maximum (FWHM) of 55.6\AA. The final stacked image has a delivered image quality of 1.2\arcsec\ and a limiting magnitude $m_{o3,5\sigma}=25.2$ measured in a 2\arcsec\ diameter aperture.

To create a multi-wavelength photometric catalog, \citetalias{shi19b} first homogenized the point spread functions (PSFs) of the `T0007' version CFHTLS broadband images to match the seeing of the $o3$ image. A Moffat profile was assumed to fit the PSF of the $o3$ image with the seeing-dependent parameter $\beta=3.0$ and FWHM of 1.2\arcsec. By running the code \texttt{IDL\_ENTROPY}, \citetalias{shi19b} derived the convolution kernel of each broadband image and convolved it accordingly. We run the \texttt{SEXTRACTOR} software \citep{sextractor} using the dual-image mode with the $o3$ image as the detection band while measuring the photomerty in the $ugri$ bands.

Based on our multi-wavelength catalog, LAEs are selected using the following criteria:
\begin{eqnarray}\label{lae_color}
o3-g<-0.9~~\wedge~~{\rm S/N}(o3)\geq 7 ~~\nonumber \\
\wedge~~[~u-g>1.2~~\vee~~{\rm S/N}(u)<2 ~]
\end{eqnarray}
The $o3-g$ color  ensures the rest-frame equivalent width $W_0 \gtrsim 20$\AA. The $u-g$ color criterion or non-detection in the $u$ band ensure that the colors of these sources are consistent with being at $z \gtrsim 2.7$.

In total, 93 LAEs are identified over an effective area of 1,156 arcmin$^2$. Of these, none has an X-ray counterpart in the existing XMM data\footnote{The flux limits of the XMM data in  the D1 field are $2.5\times 10^{-15}$ erg~cm$^{-2}$~s$^{-1}$ in the 0.5--2 keV band, and $2 \times 10^{-14}$ erg~cm$^{-2}$~s$^{-1}$ in the 2--10 keV band.} \citep{chiappetti05}, suggesting that AGN contamination is likely low; however, we cannot rule out X-ray faint AGN in our sample. The brightest source in the sample, labelled as QSO30046 in the \citetalias{shi19b} catalog, is  classified as an active galactic nuclei (AGN) at $z_{\rm spec}=3.86$ in the VIMOS VLT Deep Survey \citep[][]{vvds}. The blue $o3-g$ color is likely a result of  Ly$\beta$ and O~{\sc iv} emission falling into the $o3$ band. QSO30046 is removed from the LAE catalog. The contamination from low-redshift line emitters is expected to be low because the adopted $o3-g$ color cut corresponds to the observer-frame line equivalent width of $\approx$80\AA, and thus should be conservative enough to exclude most  [O~{\sc ii}] emitters ($W_0 \lesssim 60$\AA) at $z=0.34$ \citep{hogg98,ciardullo13}.

\subsection{Protocluster and Field LAEs}\label{subsec:protocluster}
Among the 93 LAEs, \citetalias{shi19b} reported that 24 sources reside in a significant LAE overdensity near a spectroscopically confirmed protocluster at $z=3.13$ \citep{toshikawa16}. The observed overdensity parameter  is comparable to that measured for several confirmed protoclusters \citep[e.g.,][]{kurk00,venemans07,lee14,dey16}. Moreover, a  luminous and extended Ly$\alpha$ nebula ($L_{{\rm Ly}\alpha} \approx 2\times 10^{43}$~erg~s$^{-1}$, with a half-light radius of 40~kpc) was also discovered  near the overdensity, as is the case for several known protoclusters \citep{steidel00,matsuda04,badescu17}. These findings lend credence to the possibility that these LAEs trace a protocluster structure at $z=3.13$ which will evolve into a galaxy cluster with a total mass of $(1.0-1.5)\times 10^{15}M_\odot$ by $z=0$.

We use the the protocluster LAE sample defined by \citetalias{shi19b}. Briefly, \citetalias{shi19b} constructed the LAE surface density map by smoothing the LAE positions by a Gaussian kernel with the FWHM of 10~Mpc (5.1\arcmin). The highest  overdensity region outlined by an isodensity contour contains 21 LAEs within an area 72.8~arcmin$^2$ corresponding to a surface overdensity $\delta_\Sigma = 3.3\pm 0.9$ ($\delta_\Sigma \equiv (\Sigma-\bar{\Sigma})/\bar{\Sigma}$) where $\bar{\Sigma}$ and $\Sigma$ are mean and local surface overdensities. Additionally, we include three  sources which reside in an LBG overdensity  and were later spectroscopically confirmed by \citet{toshikawa16} to lie at $z=3.13$; they are also recovered by our LAE selection. Thus, our protocluster LAE sample consists of 24 members; the remainder are referred to as `field LAEs'. 
While these definitions are simplistic, they provide us with a rare opportunity to conduct quantitative and fair comparison of the two environmental subsamples. 
For more details of the protocluster, its galaxy members, and their distribution, we refer interested readers to \citetalias{shi19b} (their Section~4.1 and Figure~6).

\subsection{Ancillary Data}\label{subsec:vandels}

We make use of the VANDELS public spectroscopic survey DR2 data \citep{mclure18,vandels}. The majority of VANDELS targets are drawn from three categories: bright star-forming galaxies ($i_{AB}\leq 25$) at $z=2.4-5.5$; massive quiescent galaxies at $z=1.0-2.5$ with $H_{AB} \leq 22.5$; and faint star-forming galaxies at higher redshift range ($25 \leq H_{AB} \leq 27$, $i_{AB} \leq 27.5$ at  $z=3-7$). To  the combined galaxy sample, we apply further selection criteria on the VANDELS galaxies to construct an LAE-like sample, which include: 
 1) the  redshift quality flag value is 3, 4, or 9 yielding $\gtrsim 80$\% confidence in redshift determination; 2) Ly$\alpha$ line in emission is visible in the 1D spectrum. Our selection results in 18 galaxies at $z=3.0-4.9$.

Starting from the publicly available 1D spectra, we compute Ly$\alpha$ luminosities by integrating the flux density over a wavelength window of 50\AA, centered at the redshifted Ly$\alpha$ emission ($\lambda_{\rm{Ly}\alpha}=1215.67(1+z)$\AA). 
The continuum flux density in the window is determined by the interpolation from the flux density blueward and redward to the window, which is then subtracted to obtain the line flux\footnote{We do not correct for the Ly$\alpha$ flux falling outside the slitlet and therefore the Ly$\alpha$ luminosity is likely underestimated.}.

The power-law slope of the UV spectrum $\beta$ (where $f_{\lambda} \propto \lambda^{\beta}$) and UV continuum luminosity at rest-frame wavelength 1700\AA\ ($L_{1700}$) are determined from the VANDELS multi-wavelength photometric catalog. For each galaxy, $\beta$ is computed through the linear fitting of the flux densities of the bands which sample the UV portion of the SED but not Ly$\alpha$: i.e., the $izH$ bands at $z>3.2$ and the $iz$ bands at $z\leq 3.2$. 
Once $\beta$ is determined, $L_{1700}$ is  estimated by interpolating the $i$ band flux density. The rest-frame Ly$\alpha$ equivalent width ($W_0$) is derived from the ratio of Ly$\alpha$ flux to the continuum flux density at the same wavelength. The rest-frame $W_0$ ranges from 0.4\AA\ to 440\AA, with a median of 16\AA. \\

 \section{Characterizing the extended Lyman Alpha Emission}\label{sec:lah}

\begin{figure*}
\epsscale{1.2}
\plotone{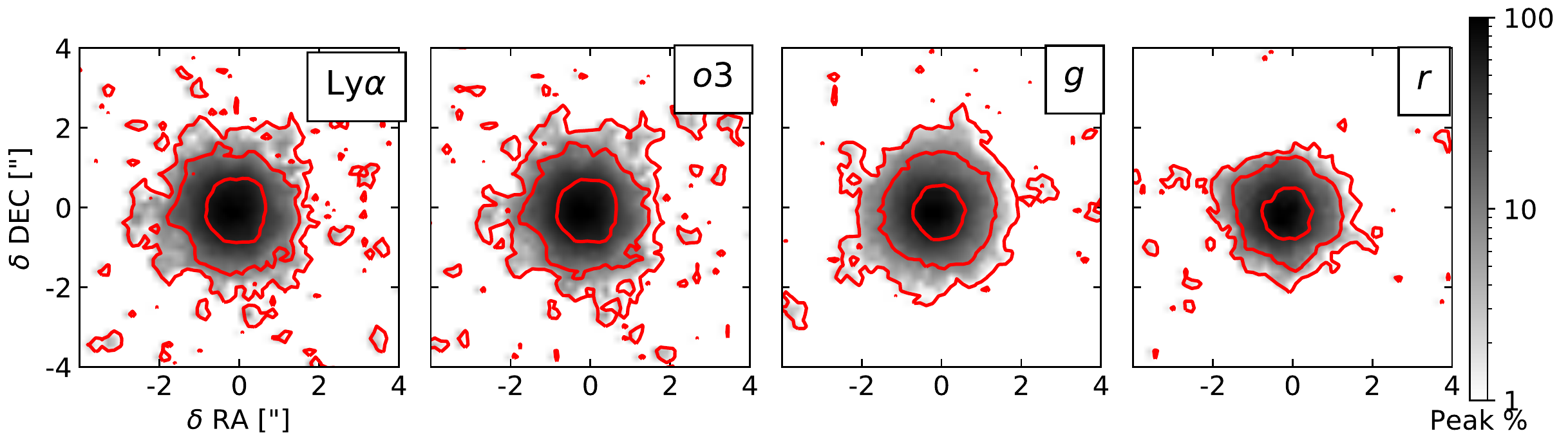}
\caption{
  Stacked images of LAEs in the D1 field. The Ly$\alpha$ image is constructed from the $o3$ and $g$ band data as described in Section~\ref{sec:lah}; both $o3$ and $g$ passbands sample near $\lambda_{\rm{rest}} \approx 1216$ \AA, and thus include the line and continuum emission to a varying degree. The $r$ band samples the UV continuum at $\lambda_{\rm{rest}} \approx 1700$\AA. In each panel, the red contours show the positions at which the surface brightness falls to 50\%, 10\%, and 1\% of the peak  brightness.}
\label{contours}
\end{figure*}

We investigate their average properties utilizing image stacking analyses. The details of our adopted procedure are given in \citet[][X17, hereafter]{xue17}. 
We create a Ly$\alpha$ line image by subtracting the continuum flux from the $o3$ image. The Ly$\alpha$ flux is determined as $F_{\rm Ly\alpha}=af_{\nu, o3}-bf_{\nu, g}$ where the line flux, $F_{\rm Ly\alpha}$, and the monochromatic flux density, $f_\nu$ are given in units of erg s$^{-1}$ cm$^{-2}$ and erg s$^{-1}$ cm$^{-2}$ Hz$^{-1}$, respectively. $a$ and $b$ are coefficients given in units of Hz, which capture the optical depth of the intergalactic medium \citep{inoue14}, which depend on the filter width, redshift, and the UV spectral shape. The detail is given in the Appendix~A of \citetalias{xue17}. 
For a $z=3.1$ galaxy with a flat UV spectrum (i.e., $\beta=-2$), $a=7.3\times 10^{12}$ Hz and $b=7.7\times 10^{12}$ Hz.

We perform image stacking on the PSF-homogenized images in the $o3$, $g$, $r$, and Ly$\alpha$ bands. For each LAE, a $1' \times 1'$ cutout is created centered on the galaxy. 
We run \texttt{SEXTRACTOR} for detection and mask out the off-center sources using the output segmentation map.
We visually inspect the cutout images and find that 11 LAEs in our sample have companions within 3\arcsec, which can potentially undermine our ability to robustly estimate and remove the local sky background, a step essential to detecting low surface brightness features around galaxies. Additional six sources are clearly extended (with semi-major axis $> 2 \arcsec $) and elongated. Of these 17 sources, 4 are in the protocluster region and the remainder belong to the field. After removing them, we perform a pixel-to-pixel median image stacking on the final sample of 76 LAEs. Median stacking ensures that our results are robust against being biased by a few high-luminosity outliers. 

The sky background in the stacked image is estimated from an annular region of $[10\arcsec, 20\arcsec]$ from the center and is subtracted to produce a zero-sky image. The inner radius of the sky annulus is chosen not to include any diffuse emission from the galaxy.\footnote{At $z=3.13$, 10\arcsec\ corresponds to 78.4~kpc; even the most extended Ly$\alpha$ emission detected around normal star-forming galaxies are found to be $\lesssim 25$~kpc, in their exponential scale-lengths \citep[e.g.,][]{steidel10}.} Using this method, we create the image stacks for our full LAE sample.

In Figure~\ref{contours}, we show the resultant stacked images of the full LAE sample in the Ly$\alpha$, $o3$, $g$, and $r$ bands. All are normalized at their peak brightness and the contours mark  1, 10, and 50\% of the peak brightness. In the bands which primarily sample Ly$\alpha$ emission (Ly$\alpha$ and $o3$), the emission is clearly  more extended  than in the $r$ band. %
The fact that the two inner contours in the $g$-band are similar to those in the $r$-band is not surprising because the continuum flux is expected to make a larger contribution to the flux density than the line flux at the range of the observed line equivalent width. The outer contour ($\gtrsim 2$\arcsec\ from the center) closely mirrors those of the Ly$\alpha$ and $o3$ band as the contribution of the Ly$\alpha$ flux increases and the continuum flux falls off rapidly.

\begin{figure*}
\epsscale{1.0}
\plotone{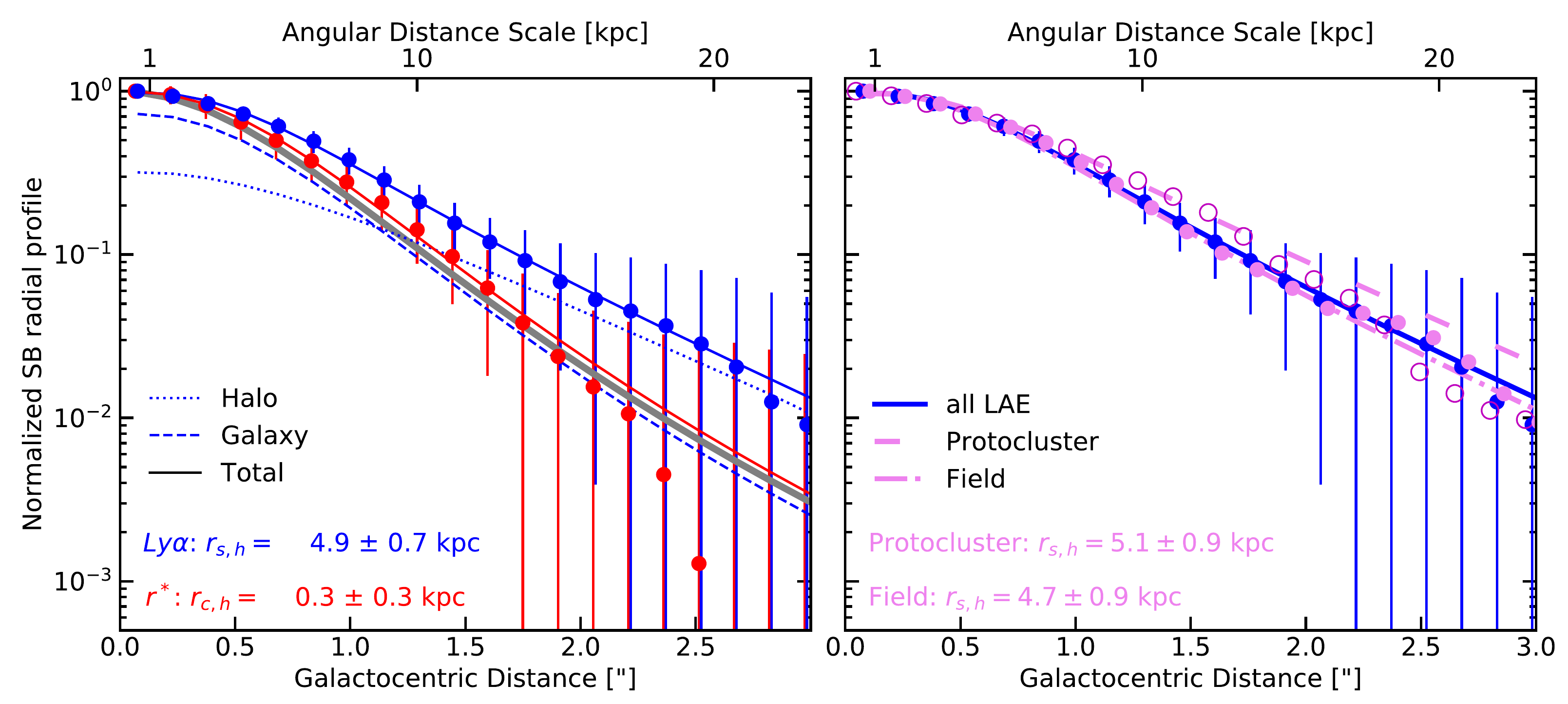}
\caption{
  \textit{Left:} normalized radial profiles of the stacked images of our LAE sample are shown in the Ly$\alpha$ (blue) and the $r$ band (red). Solid, dashed, and dotted lines of the same color show the best fit exponential profiles for the total, galaxy-, and LAH components, respectively. The exponential scale-lengths are indicated at bottom left corner. The thick grey curve shows the image PSF. Dashed and dotted curves are the continuum and halo components.
  \textit{Right:} We compare the radial profile of Ly$\alpha$ emission for the full sample (blue) with those for the protocluster (pink open circles) and field subsamples (pink filled circles). For clarity, we do not show the uncertainties of the subsamples, which are $\approx$95\% and $\approx$28\% larger in the protocluster and field subsamples than those of the full sample. The LAH sizes in protocluster- and field LAEs are indistinguishable. 
  }
\label{rad77}
\end{figure*}

 \subsection{Modeling the Surface Brightness Profile of LAH}\label{subsec:lah_size}

We measure the physical size of the Ly$\alpha$ and continuum emission from the stacked images. Because the $r$ band only samples the continuum emission, the 1D surface brightness profile of the $r$ band is approximated as an exponentially declining function:
\begin{equation}\label{dec}
I_{\rm obs} (r)={\rm PSF} \ast I_c \exp(-r/r_{\rm s,c})
\end{equation}                              
where the $r_{\rm s,c}$ is the exponential scale length of the galaxy's continuum emission, $I_c$ is the normalized surface brightness, and the symbol $\ast$ denotes the convolution. We use the PYTHON function \texttt{scipy.optimize.curve\_fit} and obtain the best fit scale length is $0.3 \pm 0.3$ kpc: this is consistent with our expectation that our LAEs are unresolved in the $r$-band. The error accounts for the full variance of the parameter space. 

The intrinsic Ly$\alpha$ surface brightness is modeled as a superposition of two exponential functions, representing the emission from the galaxy and the extended halo: 
\begin{equation}\label{eq1}
S_{\rm Ly\alpha, intrinsic}=S_c\exp(-r/r_{\rm s,c})+S_h\exp(-r/r_{\rm  s,h});
\end{equation}
The observed Ly$\alpha$ surface brightness is expressed as:
\begin{equation}\label{eq2}
S_{\rm Ly\alpha, obs}(r)={\rm PSF} \ast S_{\rm Ly\alpha, intrinsic}
\end{equation}
In Equation~\ref{eq1},  $r_{\rm s,h}$ is the scale length of the halo component; $S_c$ and $S_h$ normalize the galaxy and halo component, respectively. We determine the best-fit scale length $r_{\rm s,h}$ while fixing the $r_{\rm s,c}$ value measured in the $r$ band and obtain $r_{\rm s,h}  = 4.9 \pm 0.7$ kpc. 

In the left panel of Figure~\ref{rad77}, we show the azimuthally averaged radial profiles of the Ly$\alpha$ (blue) and $r$ band (red)  together with the best-fit exponential models as solid lines. For the ease of comparison, the profiles are normalized at the innermost angular bin. For the Ly$\alpha$ band, we also show both galaxy and LAH components as a dashed and a dotted line, respectively. The image PSF is indicated by a thick grey line.
The uncertainties include both Poisson and bootstrap sampling errors.

In order to explore how the LAH size changes with galaxies' large-scale environment, we repeat the same procedure  for the two subsamples. In Figure~\ref{rad77} (right), these measurements are shown as pink circles and lines. The uncertainty in the protocluster subsample is $\approx${95}\% ($\approx$ 28\% in the field) larger than those of the full sample as the number of galaxies in each sample is smaller. When normalized, these measures are consistent with our result for the full sample within uncertainties.  The LAH scale-lengths of  $r_{s,h}=5.1\pm 0.9$ kpc and $r_{s,h}=4.7 \pm 0.9$ kpc for the protocluster- and field subsample, respectively. Our analysis suggests that the large-scale environment is not a major determinant of LAH sizes, in agreement with the finding of \citetalias{xue17}.

\begin{figure*}
\epsscale{1.}
\plotone{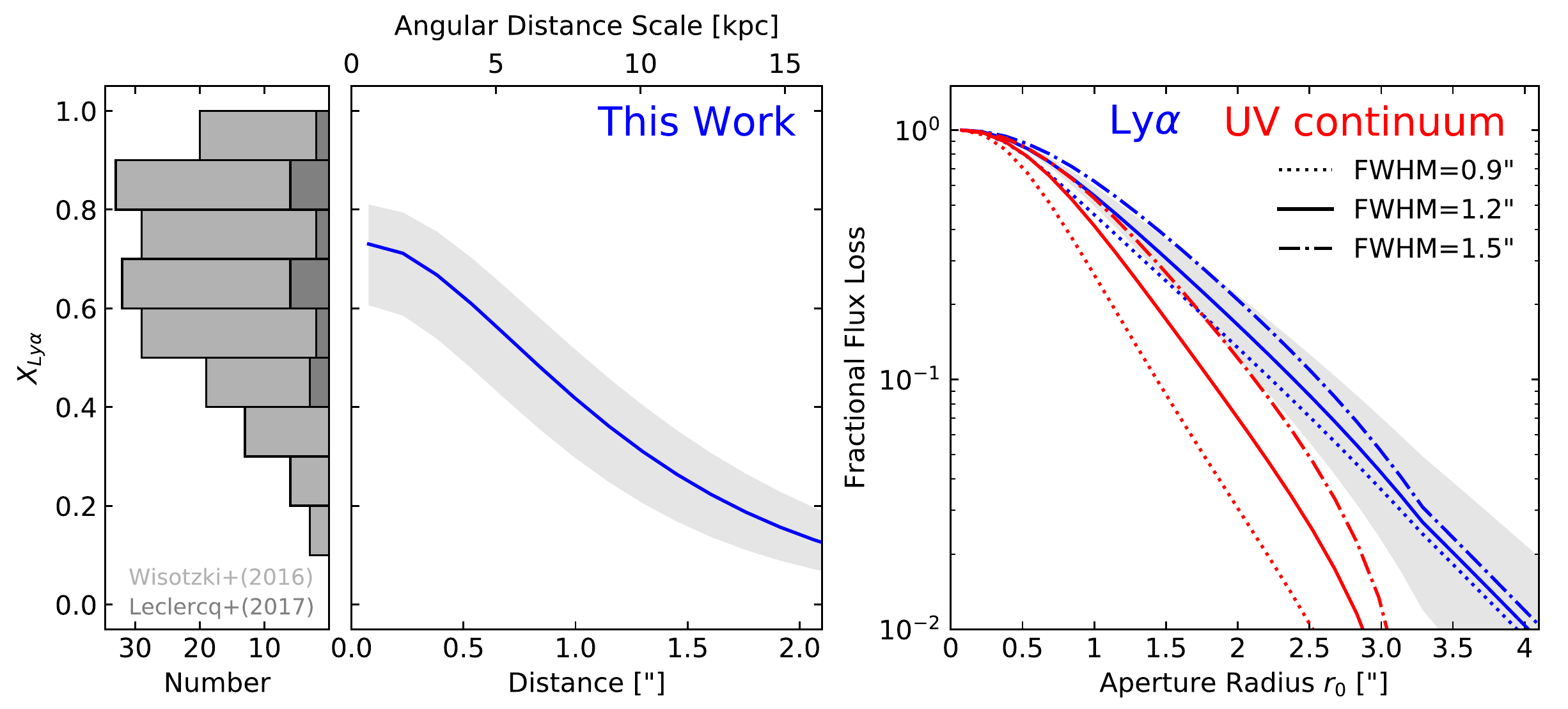}
\caption{
{\it Left:} the number distribution of the fractional contribution of the LAH to the total Ly$\alpha$ flux, $X_{\rm Ly\alpha}$, as reported by \citet[][light grey]{wisotzki}  and  \citet[][dark grey]{lecle17}.
{\it Middle:} $X_{{\rm Ly}\alpha}$ as a function of the distance from the galactic center to the inner `edge' of the LAH ($r_{\rm halo}$) computed 
using the Equations~\ref{eq3}-\ref{eq5}. The blue line indicates the best-fit model with the grey shade shows the 1$\sigma$ uncertainties.
{\it Right:} both the intrinsic size of the emission and the image PSF lead to the loss of flux falling outside a given aperture. Using our best-fit profiles, we compute the fractional loss of the UV continuum (red) and Ly$\alpha$ emission (blue) of typical LAEs as a function of aperture size assuming three representative seeing values (full-width-at-half-maximum of 0.9\arcsec, 1.2\arcsec, and 1.5\arcsec shown as dotted, solid, and dashed-dot lines, respectively). For example, at seeing 1.2\arcsec, $\approx$ 30\% (20\%) of the total Ly$\alpha$ (UV continuum) flux falls outside the 1.5\arcsec\ radius circular aperture. 
}
\label{fraction}
\end{figure*}

 \subsection{Ly$\alpha$ Halo Fraction and Aperture Correction}\label{subsec:ap_corr}
Having separated the total Ly$\alpha$ emission into the halo and galaxy components, we estimate the fractional contribution of the LAH to the total line flux. The halo fraction is defined as:
\begin{equation}\label{eq3}
X_{\rm Ly\alpha}\equiv F_{\rm h}/F_{\rm tot}
\end{equation}
where the halo and total fluxes, $F_{\rm h}$ and $F_{\rm tot}$, are computed as:
\begin{equation}\label{eq4}
F_{\rm h}=\int_{r_{\rm halo}}^{\infty} 2\pi r S_h \exp(-r/r_{\rm s,h}) dr;
\end{equation}
\begin{equation}\label{eq5}
F_{\rm tot}=\int_{0}^{\infty} 2\pi r [S_c \exp(-r/r_{\rm s,c})+S_h \exp(-r/r_{\rm s,h})] dr
\end{equation}
We note that, unlike $F_{\rm tot}$, the integration interval for $F_{\rm h}$ in Equation~\ref{eq4} is set to [$r_{\rm halo}$,$\infty$]; i.e., $F_{\rm h}$ encloses the observed flux outside the central region  where  $r= r_{\rm halo}$ serves as the inner `edge' of the LAH, or equivalently, the outer edge of the galaxy.
This definition is motivated by the fact that, at a distance comparable to or smaller than the size of the host galaxy, the physical meaning of the LAH (as a component separate from the galaxy emission) becomes ambiguous in all but a mathematical sense. However, our definition of $F_{\rm h}$ would be identical to that used in \citet{wisotzki} and \citet{lecle17} if $r_{\rm halo}$ is set to zero.

In the middle panel of Figure~\ref{fraction}, we show  $X_{{\rm Ly}\alpha}$ as a function of $r_{\rm halo}$.
The $1\sigma$ uncertainties (grey shade) are computed by bootstrapping the radial profile measurements. At $r_{\rm halo}=0$, our halo fraction is $73^{+8}_{-12}$\%, in good agreement with those estimated by \citet{wisotzki} and \citet{lecle17} (light grey and dark grey histograms, respectively, in the left panel of Figure~\ref{fraction}), and also consistent with the expectation from simulations with a realistic treatment of the ISM \citep{verhamme12}. When evaluated at $r_{\rm halo}=0.5$\arcsec\ (3.9~kpc at $z=3.13$) which lies safely outside typical galaxy sizes,  the halo fraction decreases to $61^{+9}_{-12}$\%. Our analysis indicates that, for LAEs, the flux from the Ly$\alpha$ halo easily dominates over that originating from the galaxy itself, even in the most conservative estimate.

From the observational viewpoint, it is often useful to know what fraction of the flux falls outside a given aperture. The calculation is similar to that shown in Equation~\ref{eq4} except that we integrate the {\it observed} surface brightness  instead of the intrinsic one. In the right panel of Figure~\ref{fraction}, we show the fractional  loss of the Ly$\alpha$ (purple) and continuum $r$-band (blue) flux as a function of aperture radius, $r_0$. 
We start by convolving the best-fit radial profiles with the gaussian kernels with the full-width-at-half-maximum of 0.9\arcsec\ (dotted), 1.2\arcsec\ (solid), and 1.5\arcsec\ (dash-dotted), respectively, to simulate the realistic range of seeing in ground-based observations, which also affect the flux loss.

In  a 3\arcsec\ aperture in diameter ($r_0=1.5$\arcsec), the flux loss is $32\% \pm 6$\% in the Ly$\alpha$ image, much larger than $19.0\% \pm 0.4$\% in the $r$ band. The aperture radius which encloses 90\% of the total flux,  $r_{90}$,  
is $2.4\arcsec \pm 0.3\arcsec$ for the  Ly$\alpha$ band, considerably smaller than $r_{90}=3.7 \arcsec \pm 0.2 \arcsec$ measured for a sample of galaxies with both H$\alpha$ and Ly$\alpha$ emission \cite{matthee16}. The discrepancy is too large to be explained by the difference in image quality.

\begin{figure*}
\epsscale{1.2}
\plotone{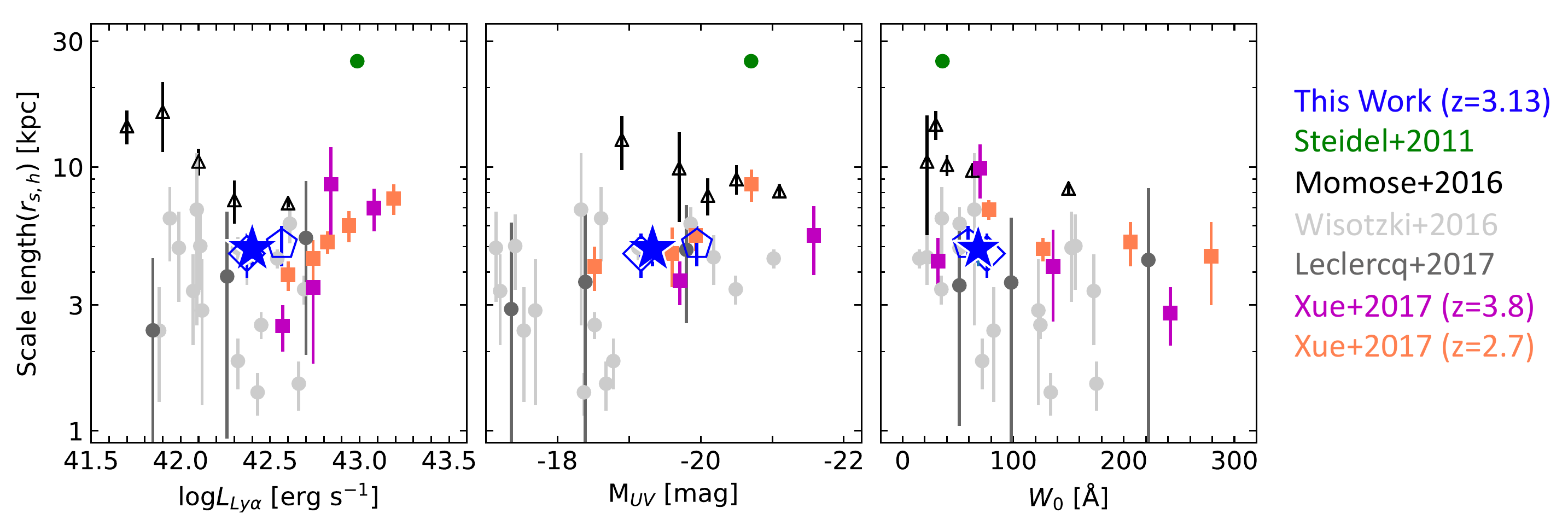}
\caption{
 LAH sizes as a function of Ly$\alpha$  luminosity, UV luminosity, and Ly$\alpha$ rest-frame equivalent wavelength. Our measurement is shown as blue stars together with the literature measurements including those from \citet[][green circles]{steidel11}, \citet[][black open triangles]{momose16}, \citet[][light grey filled circles]{wisotzki}, \citetalias{xue17} (orange filled squares for $z=3.8$ and purple squares for $z=2.7$), and \citet[][dark grey filled circles]{lecle17}. The \citet{lecle17} data are binned and the median value and standard deviation in each bin are displayed. We show the stacked LAH measurements respectively for the LAEs in the protocluster (blue open pentagon) and in the field (blue open diamond).}
\label{compare}
\end{figure*}
The origin of this disagreement is likely intrinsic.  Although the galaxies in both samples are both detected with significant Ly$\alpha$ line excess, the H$\alpha$-Ly$\alpha$  emitters tend to have larger stellar masses and  higher star formation rates (SFRs) than our LAEs. \citet{matthee16} made the stellar mass estimates based on the spectral energy distributing (SED) fitting and reported $\log{(M_{\rm star}/M_\odot)}=8.6-11.1$ with a median of 10.3 (see their Table~1).
Although we do not have a direct estimate of stellar mass for our LAE sample, photometrically selected LAEs tend to be low-mass galaxies in the range $(10^8-10^9)M_\odot$ \citep{gawiser06, nilsson11, guaita11,shi19a}. As for the SFR, \cite{matthee16} reported the range $(3-50)~M_\odot~{\rm{yr}^{-1}}$  with a median of 20~$M_\odot~{\rm{yr}^{-1}}$, once again, much larger than $\approx 5~M_\odot~{\rm{yr}^{-1}}$ for our sample by using dust-corrected SFRs generated by observed UV luminosity \citep[][also see Section~\ref{subsec:var}]{shi19a}.  
In Section~\ref{subsec:lah_origin}, we  discuss how the LAH sizes change with physical properties.

 \subsection{On the LAH sizes}\label{subsec:lah_origin}

In Figure~\ref{compare}, we show the compilation of Ly$\alpha$ size measurements in the literature as a function of line and continuum luminosity and $W_0$. The measurements of the full LAE sample, protocluster- and field subsamples are indicated as a blue star, open pentagon and open diamond, respectively. Overlaid are similar measurements based on image stacking for the photometrically selected LAEs at $z\sim2.7$ and $\sim3.8$ (\citetalias{xue17}, orange and purple, respectively) and at $z\sim2.2$ \citep[][open triangles]{momose16}. 
We also  show individual measurements for $z\sim 3-6$ star-forming galaxies from deep MUSE observations from \citet[][light grey circles]{wisotzki} and \citet[][dark grey circles]{lecle17}. The latter measurements are binned\footnote{As for  the  binning, we used  $<10^{42}$, $10^{42}-10^{42.5}$, and $>10^{42.5}$~erg~s$^{-1}$  for the line luminosity, $>-18$, $-(18-19)$, and $\leq -19$~mag for the UV absolute magnitude, $<75$, 75--150, and $\geq 150$~\AA\ for $W_0$.} for clarity. A green circle in each panel represents the stacked measurement of UV-selected star-forming galaxies  \citep[][]{steidel11}, with  a median equivalent width of 0.9\AA\ (and the range $-27\leq W_0 \leq 89$\AA), the majority of which would not  be classified as LAEs.
The remainder of the literature samples shown in Figure~\ref{compare} employed the classical definition of LAEs, i.e., $W_0\geq 20$\AA\ with the exception of the $z\sim2.7$ sample of \citetalias{xue17}, which used $W_0\geq 50$\AA.

While limited, our size measurements are in line with the existing data. In particular, the agreement is excellent when compared with the measurements that employed a similar decomposition method which simultaneously fits the galaxy and the LAH component (see Section~\ref{subsec:lah_size}), namely, those in \citet{lecle17} and \citetalias{xue17}. Disagreement with the \citet{momose14} and \citet{steidel11} may be in part due to the difference in the fitting method in that they only considered a single component. For an in-depth discussion of how the size measurement can be affected by the fitting method and image point spread function, we refer interested readers to the Appendix~C of \citetalias{xue17}.

Figure~\ref{compare} showcases the overall trend that the LAH sizes increase with both Ly$\alpha$ luminosity and UV luminosity while showing little correlation with galaxy environment. 
The very large LAH size \citep[$\approx 25$~kpc:][]{steidel11} measured for UV continuum selected star-forming galaxies lies far above these trends, suggesting that a different scaling relation may apply to non-LAEs. The larger LAH sizes for more UV-luminous galaxies imply that the flux loss we estimate in Section~\ref{subsec:ap_corr} is only applicable to typical LAEs (which are UV-faint, $M_{\rm UV}\gtrsim -20$) and cannot be generalized to all star-forming population. 

Discerning how LAH sizes and Ly$\alpha$ surface brightness profiles change with the properties of host galaxies and their large-scale structure can place strong constraints on the dominant physical mechanism that powers the Ly$\alpha$ halo (see, e.g., \citealt{matsuda12,momose16}; \citetalias{xue17}; \citealt{lecle17}).  Possible scenarios include resonant scattering  of the Ly$\alpha$ photons originating from star formation in the  CGM \citep[][]{laursen07, dijkstra09, verhamme12}, gravitational cooling radiation  \citep[][]{haiman00,fardal01,dijkstra09,lake15}, and star formation in ultra-faint satellite galaxies \citep[][]{zheng11,lake15}. While the limited nature of our measurement (based on a single stack)  prevents us from placing a new meaningful constraint on these scenarios, we stress that the mild disagreement between existing measurements, the large scatter observed in the individual measurements and the apparent dichotomy between LAEs and non-LAEs in LAH sizes seen in Figure~\ref{compare}  highlight the incompleteness of the current observational picture. Larger samples spanning a wider range of parameter space (in luminosities and large-scale environment) will be crucial in establishing clearer trends and in discriminating different physical scenarios. \\

\section{Lyman $\alpha$ Escape fraction} \label{sec:esc}

The escape fraction of Ly$\alpha$ photons is expected to be a sensitive function of not only a galaxy's dust content but also of the distribution of gas and dust therein. In a medium in which the H~{\sc i} gas and dust are uniformly mixed, resonant scattering of Ly$\alpha$ photons renders them to suffer a higher degree of extinction relative to continuum photons at a similar wavelength. As a result, a galaxy selection based on Ly$\alpha$ line equivalent width would be heavily biased towards galaxies with little to no dust. 
While such an expectation is in line with a  majority of LAEs \citep{cowie98,steidel00,gawiser06,nilsson09,guaita11}, some LAEs are very dusty \citep[e.g.,][]{lai07,pirzkal07, lai08, nilssonm09, webb09, yuma10},
suggesting that other factors may contribute to the escape of Ly$\alpha$ photons \citep{finkelstein08,finkelstein09,scarlata09}. 

The relative distribution of gas and dust is important. \cite{scarlata09} showed that the measured Ly$\alpha$-to-H$\alpha$ line ratio of local LAE analogs favors a `clumpy dust screen' scenario in which Ly$\alpha$-emitting gas is spatially segregated from the dust which exists in clumps. Clumpy dust results in a more porous medium through which Ly$\alpha$ photons can travel with more ease, thereby enhancing their transmission relative to the same amount of gas and dust in a uniform mixture.  In the `clumpy multi-phase' scenario \citep{neufeld91}, dust coexists with H~{\sc i} gas in clumps embedded in an otherwise warm, ionized medium. Such a configuration  would increase Ly$\alpha$ transmission greatly as continuum photons are more prone to dust extinction while  Ly$\alpha$ photons scatter off the clump and propagate through the ionized medium.

In this section, we present the Ly$\alpha$ escape fraction, $f_{\rm esc}$, measured for our LAE sample. We also present the {\it total} escape fraction including the contribution from diffuse Ly$\alpha$ emission, which is not accounted for in the majority of existing measurements \citep[but see][]{kusakabe18}. We also evaluate the possible correlation between $f_{\rm esc}$ and other galaxy properties and discuss possible implications of our results on the distribution of gas and dust.

\subsection{Measuring the Lyman Alpha Escape Fraction}\label{subsec:measure}

For our analyses in this and subsequent  sections, we only consider 62 LAEs for which we have robust measurements of the UV slope $\beta$. To this end, we only use the galaxies with $\Delta \beta < 0.9$.  
A majority of the sources that do not meet this criterion are simply too faint in the images and typically have inferred UV luminosities log$(L_{1700}) \lesssim 27.3$ erg s$^{-1}$ Hz$^{-1}$. 
Following the convention, we compute the Ly$\alpha$ escape fraction as:
\begin{equation}
f_{\rm esc} 
=\frac{{\rm SFR}({\rm Ly}\alpha)}{{\rm SFR}({\rm UV}) \times 10^{(0.4k_{1500}{\rm E}(B-V))}}
\label{eq6}
\end{equation}
where the color excess E($B-V$) is converted from the UV slope $\beta$ by assuming the \citet{cal2000} extinction law. For the UV- and Ly$\alpha$-based SFR, we adopt the \citet{kennicutt} calibration. $k_{1500}$ denotes the effective dust extinction at rest-frame 1500\AA.

The median (mean) value is  60~($101^{+107}_{-101}$)\% for our LAE sample. Of the 62 LAEs, nearly one third (21) have $f_{\rm esc}$ values greater than unity, nine of which lie within the protocluster (see Section~ \ref{subsec:protocluster}). 9 LAEs (2 in the protocluster) do so at a $\geq 1.5\sigma$ level. In comparison, \citet{blanc2011} reported 24\% for spectroscopically selected LAEs with \fesc\ greater than unity at $z$=1.9--3.8. Both numbers are based on dust-corrected UV continuum to infer SFRs and thus are subject to similar systematics. 
Excluding the protocluster LAEs brings down the fraction of LAEs  with $f_{\rm esc}\geq 1$ to 27\% in a  better agreement with \citet{blanc2011}. Comparison of field vs protocluster LAEs is given in Section~\ref{sec:environment}. \\

\begin{figure*}
\epsscale{1.0}
\plotone{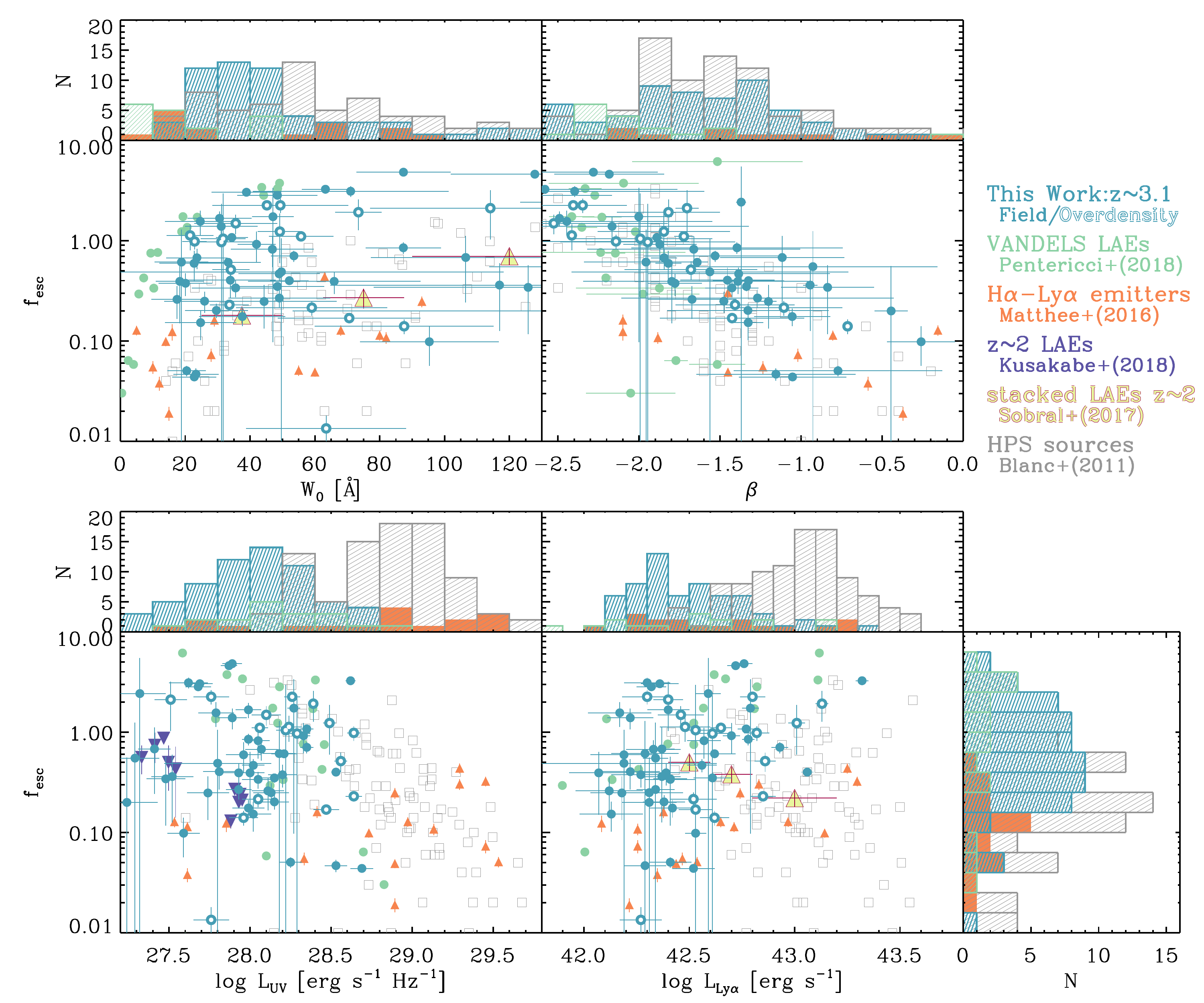}
\caption{
 Ly$\alpha$ escape fraction as a function of Ly$\alpha$ equivalent width (top left), UV continuum slope $\beta$ (top right), observed UV and Ly$\alpha$ luminosity (bottom  panels) from this work (teal circles) and literature measurements, which include the VANDELS sources \citep[green circles:][]{vandels}, $z\sim 2$ LAEs \citep[purple downward triangles:][]{kusakabe18}, the HETDEX Pilot Survey emitters \citep[grey squares:][]{blanc2011}, stacked LAEs at $z\sim 2$ \citep[yellow upward triangles:][]{soboral18}, and H$\alpha$-Ly$\alpha$ dual emitters \citep[orange upward triangles:][]{matthee16}.  Protocluster (field) LAEs from our sample are shown as open (filled) circles. The histograms  show the overall distributions of different samples in each parameter space.
   }
\label{escape}
\end{figure*}

Larger-than-unity \fesc\  values in these two samples may in part arise  from  photometric scatter, particularly in the bands used for estimating $\beta$. To quantify how robustly the estimates of $\beta$ and \fesc\ can be made for individual galaxies, we create image simulations calibrated to closely reflect the brightness and the colors of the real LAEs. While the details of our simulations are given in Appendix~\ref{appendix}, our result  suggests that the photometric recovery of $\beta$  values is  reasonably good at $\beta\approx -2$ ($\approx 0.2$~dex), as illustrated in Figure~\ref{in_out}. Photometric scatter tends to result in the recovered $\beta$ values being more positive than the intrinsic ones, leading to the underestimation -- and not overestimation -- of \fesc\ values.  Furthermore, many of our LAEs with \fesc$>1$ have relatively small $\Delta \beta$, placing their \fesc\ values  $>(2-3)\sigma$ outside the nominal $f_{\rm esc}=1$ line (see Figure~\ref{escape}). Based on these considerations, we argue that, while we cannot rule out the contribution of photometric scatter on $f_{\rm esc}>1$ sources, it is unlikely to be the sole driver of the $f_{\rm esc}>1$ sources.

There are multiple physical factors which may result in the larger-than-unity \fesc.
First, radiative transfer through the ISM of complex geometry may cause Ly$\alpha$ emission to have a preferential direction \citep[e.g.][]{verhamme12} unlike continuum emission; in this case, the ratio of  Ly$\alpha$ to UV emission would have little physical meaning. While possible, this is unlikely to be the dominant cause given the relatively tight correlation of \fesc\ and other galaxy properties (see Section~\ref{sec:dust}). Second, low-level AGN may contaminate our sample. As discussed in Section~\ref{sec:obs}, few of our LAEs have spectroscopic observations. 
It is also possible that the underlying assumptions we make about these galaxies are wrong: they have had relatively continuous and prolonged star formation histories and have solar or moderately sub-solar metallicities. While Ly$\alpha$ luminosity traces instantaneous star formation ($<10$~Myr), far-UV continuum luminosity represents the SF activity averaged over the last $\approx 100$~Myr \citep{kennicutt}. Similarly, substantially subsolar metallicities would result in higher ionizing radiation at a fixed mass. Thus, highly episodic SF activities, extremely young ages, and low metallicities will all lead to a higher Ly$\alpha$ output relative to the UV and can result in seemingly unphysical $f_{\rm esc}$ values  \citep{charlot93,malhotra02,kashikawa12,hashimoto17}.

Alternatively, the dust law we assume to convert measured UV flux density to SFR may not be appropriate. Existing studies  \citep[e.g.,][]{reddy06b,siana08} found that UV-selected star-forming galaxies dominated by younger stellar populations may be better characterized by a Small Magellanic Clouds-like dust law \citep{gordon03}.  \citet{kusakabe15} reached a similar conclusion for a large sample of LAEs. However,  an SMC-like dust law would exacerbate the problem at hand. Assuming the SMC law, a given UV slope $\beta$ would correspond to a smaller color excess E$(B-V)$, leading to an even larger $f_{\rm esc}$ than previously. 

\subsection{Total Escape Fraction of Ly$\alpha$ photons} \label{subsec:total_fesc}
After excluding the galaxies with $f_{\rm esc} \geq 1$, the median (mean) value of the Ly$\alpha$ escape fraction for our sample is 36~(40$\pm$26)\%. Our estimation is in good agreement with other LAE samples in the literature\footnote{The Ly$\alpha$ equivalent width cut is slightly different among the literature, from 20\AA\ \citep{blanc2011}, 65\AA\ \citep{oteo15},  $16$\AA\ \citep{sobral16} and $4$\AA\ for H$\alpha$-detected galaxies studied by \cite{matthee16}.}, including $29^{+40}_{-29}\%$ \citep{blanc2011}, $38\% \pm 11\%$ \citep{oteo15} and $37\% \pm 7\%$ \citep{sobral16}. One exception is the H$\alpha$-emitting LAE sample \citep[$11 \pm 11$\%:][]{matthee16}, which shows considerably lower $f_{\rm esc}$ values than the rest. As discussed in Section~\ref{subsec:ap_corr}, these H$\alpha$-Ly$\alpha$ emitters tend to be more UV-luminous and have larger stellar masses than those identified  based on Ly$\alpha$ emission alone, likely signaling that $f_{\rm esc}$ correlates with these physical parameters (see Section~\ref{subsec:var}).

So far, we have considered the $f_{\rm esc}$ values measured within the galaxy-sized apertures as has been done in the literature. 
If Ly$\alpha$ emission from the LAH originates from the same H~{\sc ii} regions as that from the galaxy, the \fesc\ estimates would need to be revised to account for the additional contribution from the LAH. 
Using the radial surface brightness profiles of the stacked Ly$\alpha$ and UV images (Section~\ref{subsec:ap_corr}, Figure~\ref{fraction}), we  estimate the total Ly$\alpha$ escape fraction by statistically correcting for the expected flux loss.
For the correction, we use the radius of the effective circular aperture, defined as $r_0=\sqrt{A_{\rm iso}/\pi}$ of each LAE. $r_0$  ranges from 0.6\arcsec--2.0\arcsec\ with a median of 1.0\arcsec~(7.8 kpc).  The Ly$\alpha$ flux loss ranges in 6--77\% with a median of 54\%, mainly due to the extended nature of the halo. While smaller, the UV flux loss due to PSF blurring (seeing 1.2\arcsec) is also not negligible, ranging in 10--69\% with a median of 40\%.

We find the median (mean) value of the {\it total} Ly$\alpha$ escape fraction, $\langle f_{\rm esc,tot}\rangle$, is 46~($51\pm 32$)\% in good agreement with $42\pm24$\% reported by \citet{kusakabe18}, which includes the LAH flux in their estimation. Similar to ours, \citet{kusakabe18}  also defined their LAEs to be galaxies with  $W_0\geq 20$\AA. While our value is higher than those in the literature, the relative increase is modest at $\approx 29\%$ when the flux loss correction is made consistently for both UV and Ly$\alpha$ fluxes. Additionally, the LAH size for typical LAEs is small enough (4.5~kpc corresponds to 0.57\arcsec\ at $z=3.13$) that galaxy-sized apertures can contain much of the flux.

\subsection{Variations of Ly$\alpha$ Escape Fraction}\label{subsec:var}

We explore how $f_{\rm esc}$ values correlate with the galaxies' UV and Ly$\alpha$ properties by combining our own measurements with those in the literature. Since most of these measures do not include the LAH component, we opt to use our $f_{\rm esc}$ values before the LAH correction. Given the modest change that it brings, our conclusion should not change substantially.

In Figure~\ref{escape}, we show the correlations of \fesc\ with $W_0$, UV slope ($\beta$), and line and continuum luminosities. In all cases, we show protocluster LAEs with larger symbols than those in the field. Other measurements include 89 galaxies from the HETDEX Pilot Survey \citep[HPS hereafter: ][]{blanc2011}, 17 LAEs with H$\alpha$ detection \citep{matthee16}, and 18 spectroscopic Ly$\alpha$ detections from VANDELS. 
The average measurements via stacking analysis of $z\sim2.2$ LAEs \citep{sobral16} are also shown.

\begin{table*}
\caption{Rank Correlation Coefficients from Kendall's $\tau$  and Spearman's $\rho$ Tests}
\begin{center}
\begin{tabular}{c|cccc}
\hline
\hline
Samples &   $W_0$  & $\beta$   & $L_{\rm UV}$ & $L_{{\rm Ly}\alpha}$ \\ 
\hline
\multicolumn{2}{c}{} & Kendall's $\tau$ Test: $\tau_{\rm K}$~(p$_{\rm K}$)\\
\hline
This Work  &  0.118 (0.174) &  0.653 ($<0.001$) & 0.046 (0.597) & 0.189 (0.030)   \\
HPS & 0.013 (0.856) &  0.704  ($<0.001$) & 0.518 ($<0.001$) & 0.120 (0.083)  \\
This Work+HPS &  0.037 (0.492) &  0.645 ($<0.001$) & 0.418 ($<0.001$) & 0.100 (0.060)  \\
This Work+HPS+VANDELS &  0.032 (0.522) &  0.625($<0.001$) & 0.403 ($<0.001$) & 0.045 (0.372)  \\
\hline
\multicolumn{2}{c}{} & Spearman's $\rho$ Test: $\rho_{\rm SR}$~(p$_{\rm SR}$) \\
\hline
This Work  &  0.169 (0.188) &  0.835 ($<0.001$) & 0.0786 (0.543) & 0.287 (0.023)  \\
HPS & 0.048 (0.636)&  0.872 ($<0.001$)& 0.656 ($<0.001$) & 0.160 (0.113) \\
This Work+HPS &  0.065 (0.410)&  0.827 ($<0.001$) & 0.566 ($<0.001$) & 0.144 (0.068) \\
This Work+HPS+VANDELS &  0.057 (0.446) &  0.807 ($<0.001$) & 0.544 ($<0.001$) & 0.069 (0.361) \\
\hline
\end{tabular}
\end{center}
\label{tab1}
\end{table*}

The HPS LAEs span a comparable range of UV slope and $W_0$ to our sample, but tend to have higher line and UV luminosities; the H$\alpha$-Ly$\alpha$ emitters studied by \citet{matthee16} are much more UV-luminous than the rest. The VANDELS sources are comparable to our LAEs in both luminosities but have considerably bluer UV slopes and smaller $W_0$.

In order to test how these physical properties affect $f_{\rm esc}$, we use two non-parametric ranked correlations, namely, the Kendall's $\tau$ and Spearman's $\rho$ rank correlation coefficients \citep[][]{kendall38,spearman04}. Both tests are run for each parameter shown in Figure~\ref{escape} using two Python scripts  \texttt{scipy.stats.kendalltau} and \texttt{scipy.stats.spearmanr}. To check the consistency, we run these tests on our dataset and the HPS dataset separately then  on the combined dataset (our LAEs+HPS and LAEs+HPS+VANDELS). We do not include the \citet{kusakabe18} measures  on our tests because they represent the stacked averages. 
In Table~\ref{tab1}, we list the correlation coefficients ($\tau$ and $\rho$) and the probabilities of null hypotheses ($p$-value).
We consider the case with either $\tau$ or $\rho$ is larger than 0.6 as a robust correlation while $\approx$0 values in these parameters indicate no correlation. 

We find a clear anti-correlation between $f_{\rm esc}$ and the UV continuum slope $\beta$, in agreement with existing studies \citep[e.g.,][]{blanc2011,song14}.  This is not surprising considering that we use the $\beta$ value as a proxy for dust reddening to correct the SFR$_{\rm UV}$ in our estimation of $f_{\rm esc}$. 
A weaker correlation with observed UV luminosity is notable: the trend is clear when the combined dataset is considered which spans two orders of magnitude in luminosities. In this parameter space, the H$\alpha$-Ly$\alpha$ emitters are no longer significant outliers but part of the correlation at the highest luminosity end.  The trend is likely related to the fact that lower-luminosity star-forming galaxies tend to have bluer $\beta$ values at high redshift \citep[e.g.,][]{bouwens12,bouwens14}. We further quantify the luminosity dependence in Section~\ref{subsec:luv_dependece}.

\begin{figure*}
\epsscale{0.9}
\plotone{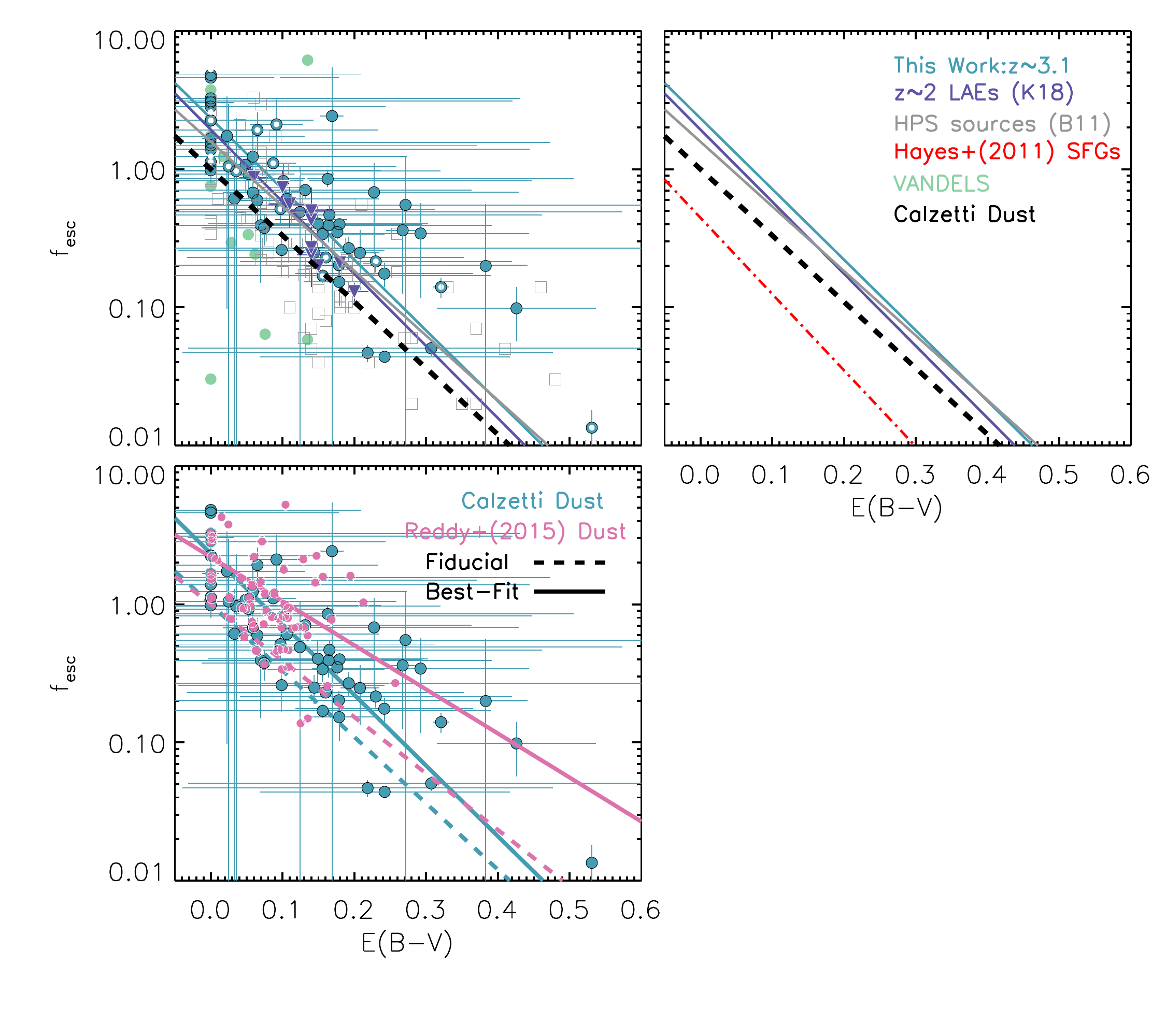}
\caption{ 
{\it Top-left:} Ly$\alpha$ escape fraction as a function of E($B-V$). Teal circles, purple downward triangles, green circles and grey open squares represent measurements of our LAEs, K18, VANDELS and HPS galaxies assuming \citet{cal2000} dust lae. Teal, purple and grey solid curves are the best models fitting to our work, K18 and HPS galaxies, respectively. Black dashed line is the expectation given the \citet{cal2000} law with no resonant scattering. {\it Top-right:} comparison among the $f_{\rm esc}$ dependence on E($B-V$) of different datasets. Red dash-dotted line is the best model for the H11 galaxies, while the remaining lines are identical to those in the top-left panel. {\it Bottom-left:} comparison between the \fesc\ measurements assuming the \citet[][teal]{cal2000} and those assuming \citet[][pink]{reddy15} dust laws of our LAEs. The best models fitting to the two measurements (solid lines) and the expectation given the \citet{cal2000} and \citet{reddy15} law (dashed lines) are also shown.
   }
\label{fig:ebv}
\end{figure*}

No correlation is  found with line luminosity or $W_0$. This is in contrast to the anti-correlation between $L_{{\rm Ly}\alpha}$ and $f_{\rm esc}$ reported by \citet{sobral16} which is based on stacking analyses. While their measurements are not inconsistent with the mean values of the combined dataset, the intrinsic scatter is  too large to discern any meaningful correlation.

All in all, we conclude that the galaxy's dust reddening and UV luminosity are two primary determinants of the escape fraction of Ly$\alpha$ photons. In the next section, we further quantify these dependencies further to shed light on the physical driver of their escape. \\

\section{The Role of Dust and UV Luminosity on \fesc}\label{sec:dust}

We examine how dust opacity affects the Ly$\alpha$ escape fraction. Following \citet{hayes11}, we express the escape fraction as:
\begin{equation}\label{2d}
f_{\rm esc}=C_{{\rm Ly}\alpha}\cdot 10^{-0.4E(B-V)k_{{\rm Ly}\alpha}}
\end{equation}
When  $C_{{\rm Ly}\alpha}=1$, $k_{{\rm Ly}\alpha}$ denotes the effective dust attenuation at Ly$\alpha$ wavelength. Even when $C_{{\rm Ly}\alpha}\neq 1$, $k_{{\rm Ly}\alpha}$ can still be thought of dust attenuation in a relative sense. As a reference, the escape fraction expected for non-resonant photons at the same wavelength is $k\approx 12$ and  $C=1$ for the \citet{cal2000} dust law. 

In Figure~\ref{fig:ebv}, we show the positions of our LAEs in the $\log f_{\rm esc}$ -- E($B-V$) space. There is a clear trend that $f_{\rm esc}$ decreases with increasing dust reddening, in qualitative agreement with existing studies at high redshift \citep{hayes11,blanc2011,song14} and at low redshift \citep{atek08,scarlata09,finkelstein11,cowie11}.

Using a Python function \texttt{scipy.optimize.curve\_fit}\footnote{The Python function \texttt{scipy.optimize.curve\_fit} uses a non-linear least squares to fit a function $f$ to a given data. The best-fit model is the one in which the sum of the squared residuals of $f(x_{\rm data})-y_{\rm data}$ is at its minimum.}, we determine the best-fit values for both parameters for our full LAE sample. In this figure, $k_{{\rm Ly}\alpha}$ and $C_{{\rm Ly}\alpha}$ appear as the slope and the intercept, respectively. By treating $C_{{\rm Ly}\alpha}$  as a free parameter, we do not require that the maximum $f_{\rm esc}$ value be unity at zero reddening. As discussed earlier, unphysically high $f_{\rm esc}$ values may result from low-metallicity or young ages. Alternatively, \citet{hayes11} argued that disparate locations from which stellar and nebular emission originate may have caused $C_{{\rm Ly}\alpha}\approx 0.45$ at zero reddening of stellar emission for $z\sim2$ galaxies. Similarly, \citet{atek14} found $C_{{\rm Ly}\alpha}$ much less than unity for $z\sim0.3$ LAEs.

We apply the same fitting procedure for i) the HPS sources; ii) the stacked averages of the $f_{\rm esc}$ values measured by \citet{kusakabe18}; and iii) our full LAE sample combined with the HPS sources. Our results are listed in Table~\ref{tab2} and indicated as various straight lines in Figure~\ref{fig:ebv}. All these measures lie well above the best-fit scaling law determined by \citet[][the orange line in the top right panel of Figure~\ref{fig:ebv}]{hayes11} at a fixed reddening.

\begin{table*}
\caption{The Dependence of Ly$\alpha$ Escape Fraction on Dust Opacity }
\begin{center}
\begin{tabular}{cccccccc}
\hline
\hline
Sample & $k_{{\rm Ly}\alpha}$ & $C_{{\rm Ly}\alpha}$ & $\chi^2$ (DoF\tablenotemark{*}) & $k_{{\rm Ly}\alpha}$ & $C_{{\rm Ly}\alpha}$ & $k_{\rm Ly \alpha}$ & $\chi^2$  (DoF\tablenotemark{*})  \\
& & Calzetti Dust & &  ($f_{\rm esc} \leq 1$) & & ($C_{\rm Ly\alpha}=1$) &  \\
\hline
{\bf This Work} & $12.79 \pm 2.31$ & $2.33 \pm 0.10$ & 31.03 (60) & $5.82 \pm 1.00$ & $0.92 \pm 0.10$ & $4.34 \pm 2.06$ & 59.16 (61) \\
\citet[][HPS]{blanc2011} & $10.02 \pm 2.00$ & $1.35 \pm 0.15$ & 26.79 (87) & $8.23 \pm 1.07$ & $0.72 \pm 0.06$ & $7.35 \pm 1.38$ & 28.66 (88) \\
\citet{kusakabe18} & $ 13.05 \pm 2.11$  & $1.93 \pm 0.40$ & -  & - & - & $7.41 \pm 1.00$ & -  \\
{\bf This Work+HPS} & $12.21 \pm 1.60$ &  $1.88 \pm 0.12$ & 65.58 (149) & $6.72 \pm 0.78$ & $0.76 \pm 0.05$ & $6.02 \pm 1.32$ & 88.68 (150) \\
\hline
c.f. \citet{hayes11} & 13.8 &0.445 & -  & - & - & -  & - \\
\hline
\end{tabular}
\tablenotetext{*}{Goodness-of-the-fit is given as the total chi-square while the degree of freedom is the number of LAEs included in the fit minus the number of free  parameters.}
\end{center}
\label{tab2}
\end{table*}

In all samples, we find that $k_{{\rm Ly}\alpha}\approx k_{1216}$: i.e., the relative attenuation scales similarly with dust reddening for both Ly$\alpha$ and continuum photons. The trend applies to all samples despite the  large spread in $C_{{\rm Ly}\alpha}$ among them.  Simultaneously, the mean Ly$\alpha$ optical depth  -- defined as $\tau_{{\rm Ly}\alpha}\equiv -2.5 \log(f_{\rm esc})/{\rm E}(B-V)$ -- is consistently lower than non-resonant photons at similar wavelength in all  but the \citet{hayes11} sample. This is visualized in the top right panel of Figure~\ref{fig:ebv} where we compare the best-fit scaling laws from the literature to the expectation from the \citet{cal2000} law at $\lambda_{\rm rest}=1216$\AA\ (a thick dashed line). Nearly all of our LAEs have $f_{\rm esc}$ values much higher than this expectation (by a factor of $\approx$2, on average: see Table~\ref{tab2}). 

To test how robust our results are against the assumed dust law, we recompute both E$(B-V)$ and $f_{\rm esc}$ values assuming the \citet{reddy15} dust law (the bottom left panel of Figure~\ref{fig:ebv}). Once again, we find  $k_{{\rm Ly}\alpha}=7.98\pm2.11$ reasonably close to  $k_{1216, {\rm Reddy}}=10.33$ for continuum photons. The mean Ly$\alpha$ optical depth is significantly lower than that expected from the \cite{cal2000} dust as driven by the large $C_{{\rm Ly} \alpha}$. We conclude that Ly$\alpha$ photons -- in all LAE samples considered here -- suffer {\it similar} interstellar attenuation to continuum photons regardless of assumed dust law. 

Finally, through realistic image simulations, we quantify how well we can constrain an underlying $f_{\rm esc}$-E$(B-V)$  scaling law through photometric measurements given the  uncertainties expected in estimating line- and continuum luminosities as well as the UV slope. While the details of our simulation are given in Appendix~\ref{appendix}, Figure~\ref{esc_ebv} illustrates that our photometric  measurements can reliably  determine an underlying scaling relation.

\subsection{Comparison with ISM models}\label{subsec:ism_models}
We consider our result in the context of several ISM models in the literature. First, in a medium composed of uniform mixtures of dust and H~{\sc i} gas, the effective attenuation of Ly$\alpha$ photons should be much higher than continuum photons \citep[e.g.,][]{charlot91} as the former traverse much longer path lengths than the latter, leading to $f_{\rm esc}$ values much lower than the continuum expectation. Our results suggest the opposite. 

Alternatively, in a ``multi-phase'' medium where dust is mostly confined within  H~{\sc i} clumps embedded in an otherwise ionized dust-free medium \citep{neufeld91}, Ly$\alpha$ photons spend most of their time traveling in the intercloud medium while occasionally scattering at cloud surfaces. Thus, \fesc\ primarily depends on cloud albedo (the likelihood of surface absorption by dust) and the number of Ly$\alpha$ scattering before their eventual escape.  The consequence is that the dependence of $f_{\rm esc}$ on total dust optical depth is mild at best \citep{hansen06,duval14}. 
In contrast, continuum photons travel through clumps and undergo more severe attenuation. The selective attenuation can enhance Ly$\alpha$ $W_0$ to, in some cases, above the case B expectation. The degree of $W_0$ enhancement increases with increasing extinction.

One possible exception would be if star formation mainly occurs deep inside cold H~{\sc i} clouds where dust resides, Ly$\alpha$ photons are decimated by dust absorption before they reach the cloud surface, leaving the role of preferential resonant scattering less consequential \citep{verhamme12}.
All in all, the strong anti-correlation between $f_{\rm esc}$ and dust reddening, the lack of LAEs with excessively high $W_0$, and the larger-than-expected \fesc\ are at odds with the multi-phase ISM.

Finally, \citet{scarlata09} considered a `clumpy dust screen' through which Ly$\alpha$ photons take the paths of the least resistance (i.e., escaping through the holes with the lowest opacity between clumps). The process allows $f_{\rm esc}$ to increase while keeping $W_0$  unchanged
without invoking preferential resonant scattering \citep[also see][]{atek14}. One defining characteristic of this model is that the relative Ly$\alpha$ enhancement scales with dust extinction, resulting in $k_{{\rm Ly}\alpha}<k_{1216}$. In Figure~\ref{fig:ebv}, the effect would manifest itself as a shallower slope. Resonant scattering would increase Ly$\alpha$ optical depth, bringing the scaling relation closer to the uniform dust screen expectation by steepening the slope $k_{{\rm Ly}\alpha}$. 

The clumpy dust screen scenario also does not align well with the existing measurements. In all three LAE samples in Table~\ref{tab2}, we consistently find $k_{{\rm Ly}\alpha}\approx k_{1216}$ within uncertainties. There is  only one exception: if we only consider sources with $f_{\rm esc}\leq 1$, the formal fit  favors a considerably lower $k_{{\rm Ly}\alpha}$ values ($\approx 6$: column 4 in Table~\ref{tab2}). A similar reanalysis of the HPS sources, however, results in a marginal change in the slope, from 10.0 to 8.2, suggesting that the dramatic change we observe in our result is caused by a small sample size. Regardless, the removal of those with \fesc$\geq 1$ without a firm physical basis is arbitrary at best. If the underlying assumptions we made in deriving the escape fraction are incorrect (as discussed in Section~\ref{subsec:measure}), the correction needed to recover the true quantities would have to be uniformly applied to all LAEs. Larger sample sizes would improve the measurements, particularly, by getting a better handle on the \fesc\ values of highly reddened LAEs and their intrinsic scatter.

 \begin{figure*}
\epsscale{0.9}
\plotone{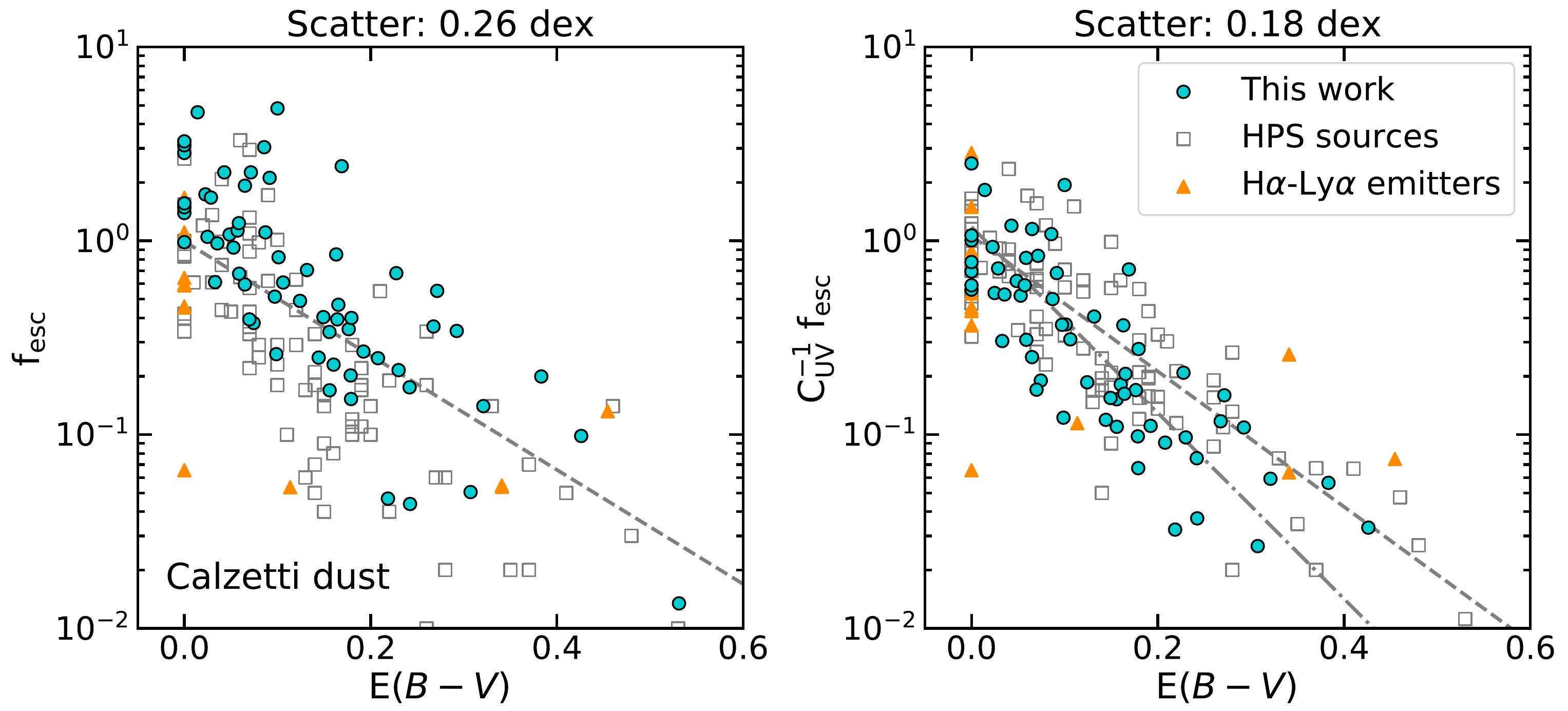}
\caption{ 
 Ly$\alpha$ escape fraction as a function of E($B-V$) before (left) and after (right) renormalizing \fesc, including measurements from this work (teal circles), HPS sources \citep[grey open squares: ][]{blanc2011} and H$\alpha$-Ly$\alpha$ emitters \citep[orange triangles: ][]{matthee16}. Dashed lines in both panels are the best-fit models assuming Equation~\ref{2d}. Dash-dotted line in the right panel shows the best-fit model assuming the same equation, while fixing the slope $k_{1216}=12$ according to \cite{cal2000} dust law. Scatters above the figure suggest the root mean square scatters between data points and the best-fit models (dashed lines). 
 }
\label{fig:fesc_ebv_corr}
\end{figure*}
In summary, our results show that Ly$\alpha$ photons undergo  interstellar dust attenuation in a similar manner to continuum radiation (i.e., $k_{{\rm Ly}\alpha}\approx k_{1216}$).  However, the overall Ly$\alpha$ transmission is roughly twice higher than that expected for continuum photons.  
The constraints derived from one of the largest compilations of LAEs to date appear to favor some form of multi-phase media conducive to the preferential transfer of Ly$\alpha$ photons. Yet, none of the models we have considered is entirely in line with our findings, hinting that the distribution of interstellar gas and dust in distant galaxies is complex. 
One possibility may be a hybrid of the two aforementioned models in which  H~{\sc i} clouds are rendered partially porous due to strong stellar feedback. This would allow for a substantial fraction of Ly$\alpha$ photons to escape from their birth cloud relatively easily and proceed to undergo selective attenuation via resonant scattering. Detailed calculations will require full radiative transfer calculations based on realistic simulations at sub-parsec scale resolution.

\subsection{Dependence of \fesc\ on UV luminosity}\label{subsec:luv_dependece}
The range of $C_{{\rm Ly}\alpha}$ changes substantially from $\approx$0.45 for the \citet{hayes11} sample to $\approx$1.4 for the HPS sources to $\approx$2 for the other LAE samples while the slope $k_{{\rm Ly}\alpha}$ remains unchanged: i.e., at a fixed reddening, \fesc\ varies up to $\approx$4 in these samples. 
The high $C_{{\rm Ly}\alpha}$ is real as the majority of our LAEs lie well above the fiducial Calzetti expectation (Figure~\ref{fig:ebv}, top right).

We evaluate the goodness-of-the-fit by computing the total chi-square value as: $\chi^2=\Sigma_i(f_{\rm esc, model}(\epsilon_i)-f_{{\rm esc},i})^2/\sigma^2_{\rm esc,i}$) where  $f_{{\rm esc},i}$ and $\sigma_{\rm esc}$ are the estimates of the Ly$\alpha$ escape fraction and its error,  $\epsilon_i$ is the estimated E$(B-V)$ value for the galaxy $i$; $f_{\rm esc,model}$ is computed using Equation~\ref{2d}.  Although the measurement errors are unlikely to obey a gaussian distribution and therefore $\chi^2$ value does not carry the usual statistical significance, it should still provide us a way to evaluate whether a given model gives a better description of the data over another.  When we refit the data while forcing $C_{{\rm Ly}\alpha}$ to be unity, we obtain unrealistically shallow slopes $k_{{\rm Ly}\alpha}$ with a much poorer agreement with the data with the total $\chi^2$ increasing from $\sim 30$ to 60. The results for our single-parameter fits are  listed in Table~2. 

The differences in $C_{{\rm Ly}\alpha}$ appear to be linked to galaxies' UV luminosities.  Figure~\ref{escape} shows that the HPS sources tend to be more UV-luminous than our LAEs by nearly an order of magnitude while the $W_0$ and $\beta$ distributions of the two samples  largely overlap. As for the H$\alpha$-Ly$\alpha$ emitters from \citet{matthee16}, which have the lowest \fesc\ values, all but four  lie at the bright end ($L_{\rm UV} > 10^{28.3}$~erg~s$^{-1}$~Hz$^{-1}$; $M_{\rm UV}<-20.7$), but they too have the $W_0$ and $L_{{\rm Ly}\alpha}$ range similar to our LAEs. Thus, we conclude that \fesc\  depends not only on the galaxy's dust content but also on its UV luminosity. 

\begin{figure*}
\epsscale{0.9}
\plotone{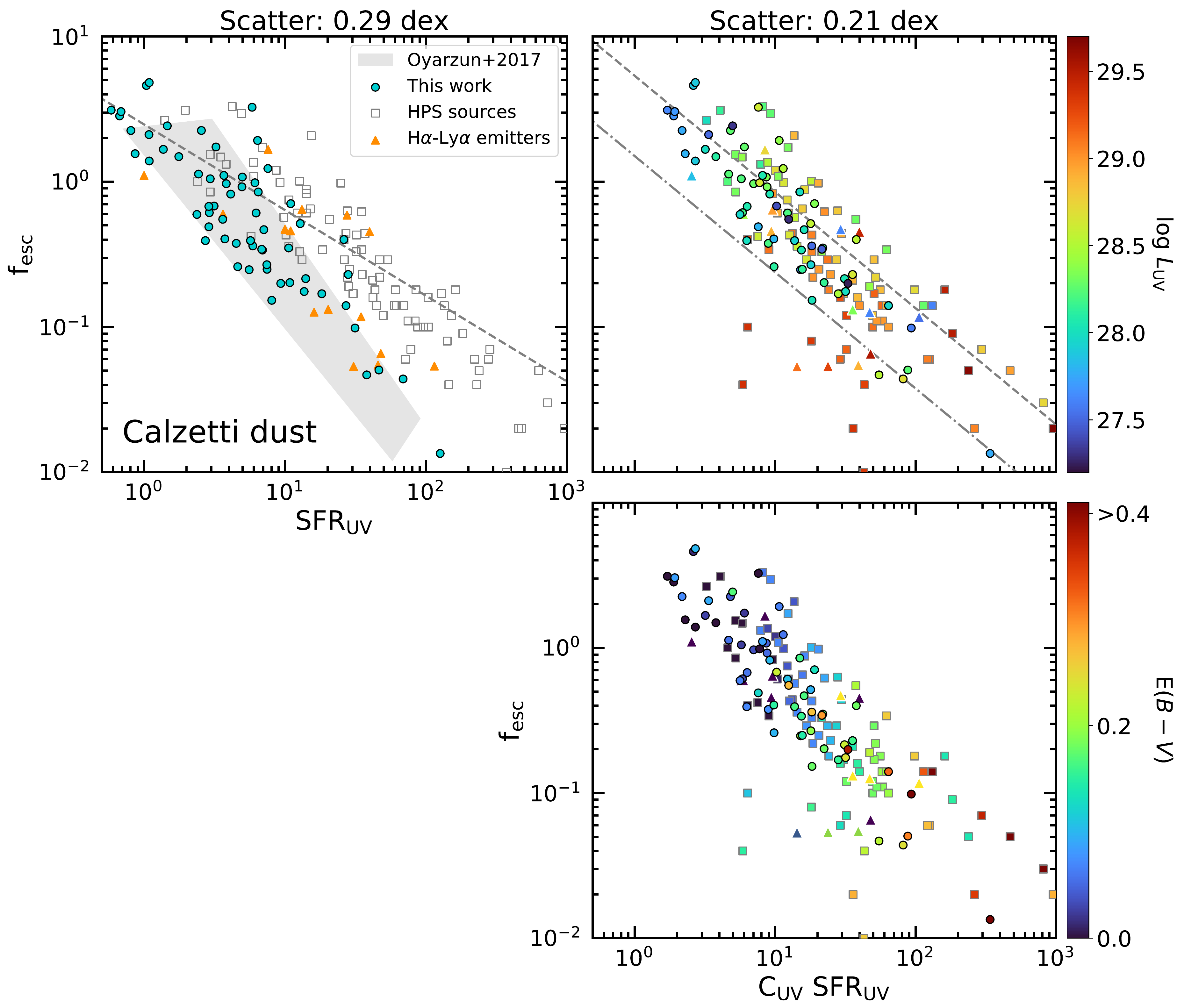}
\caption{ 
 {\it Top-left:} Ly$\alpha$ escape fraction as a function of intrinsic SFR as observed, including the measurements of this work (teal circles), HPS galaxies \citep[grey open squares:][]{blanc2011}, and H$\alpha$-Ly$\alpha$ emitters \citep[orange triangles:][]{matthee16}. Grey-shaded area in the top-left panel marks the position of \citet{oyarzun17} galaxies in parameter space. Dashed lines are the best-fit models assuming power-law correlation between \fesc\ and SFR$_{\rm UV}$. {\it Top-right:} Ly$\alpha$ escape fraction as a function of renormalized SFR$_{\rm UV}$ (C$_{\rm UV}$SFR$_{\rm UV}$). The symbol shapes are identical to those in the top-left panel, while they are color-coded by galaxies' UV luminosity. Dashed line shows the best model fitting to \fesc\ versus C$_{\rm UV}$SFR$_{\rm UV}$. All the galaxies below the dash-dotted line are identified as outliers which are excluded from calculating the rms scatter. {\it Bottom-right:} Ly$\alpha$ escape fraction as a function of renormalized SFR (C$_{\rm UV}$SFR$_{\rm UV}$, with the symbols color-coded by E($B-V$).
 }
\label{fig:fesc_sfr_corr}
\end{figure*}
Although we are limited to analyzing the Ly$\alpha$-emitting population only, our conclusions are in broad agreement with several studies of the general star-forming galaxy population:
\cite{kim20} found an anti-correlation between UV luminosity and Ly$\alpha$ escape fraction in a sample of local Lyman Break Galaxy Analogs.
\citet{stark10} found that the fraction of  Ly$\alpha$-emitting galaxies, $X$, increases with decreasing UV luminosity among the UV-selected galaxy samples\footnote{The parameter is typically denoted as $X_{{\rm Ly}\alpha}$ in the literature. Here, we denote it as $X$ to avoid confusion with another parameter defined in  Equation~\ref{eq3}. }. \citet{oyarzun17} studied a stellar-mass-selected sample of  galaxies $z$=3.0--4.5 and found that both $X$ and \fesc\ strongly anti-correlate with stellar mass \citep[see also][]{oyarzun16}. Given that UV luminosity and stellar mass broadly track each other through the star formation main sequence \citep{gonzalez11,lee12a,stark13,song16}, the latter correlation is consistent with the $L_{\rm UV}-$\fesc\ scaling we find. Similarly, \cite{weiss21} found a similar correlation between stellar mass and \fesc\ of [O~{\sc  iii}] emitters at $z=1.9-2.4$.

Motivated by the dependence of \fesc\ on UV luminosity as shown in Figures~\ref{escape}--\ref{fig:ebv}, we parametrize \fesc\ as: 
\begin{equation}\label{eq:fesc_renorm}
f_{\rm esc} = C_{\rm UV} \cdot 10^{-0.4k_{{\rm Ly}\alpha}E(B-V)}
\end{equation}
The equation is identical to Equation~\ref{2d} except for the difference that the normalization, $C_{\rm UV}$, is expressed as $C_{\rm UV}\equiv \log({\frac{L_{\rm UV}}{L_0}})^{\alpha}$ where $L_{\rm UV}$ is given in units of erg~s$^{-1}$~Hz$^{-1}$ and $\alpha$ and $L_0$ are constants. We repeat the fitting procedure using the combined sample of our LAEs and the HPS sources. The best-fit $k_{{\rm Ly}\alpha}=7.37\pm 0.98$, $\alpha=-1.62\pm 0.20$, and $L_0=(2.90\pm 1.15)\times 10^{29}$ erg~s$^{-1}$~Hz$^{-1}$. The $\alpha$ value is firmly in the negative, affirming the fact that the normalization $C$ indeed decreases with increasing luminosity.

In Figure~\ref{fig:fesc_ebv_corr}, we illustrate our result. The left panel shows  our measurements {\it as observed}, while, in the right panel, we `correct' for the luminosity dependence by showing $f_{\rm esc,renorm}\equiv C_{\rm UV}^{-1}$\fesc. Relative to the power law that best describes each set of data points (dashed lines), the scatter decreases from 0.26~dex to 0.18~dex. In short, Equation~\ref{eq:fesc_renorm} allows us to predict a galaxy's \fesc\ with a $\lesssim$50\% accuracy provided that both the UV slope $\beta$ and $L_{\rm UV}$ are  known.

The efficacy of the  luminosity-dependent \fesc\ calibration is better illustrated  in Figure~\ref{fig:fesc_sfr_corr}, where we  combine  $L_{\rm UV}$ and E$(B-V)$ into a single parameter, i.e., dust-corrected UV SFR.  In the top left panel, we once again show the measurements as observed. A clear separation between our LAEs and the HPS sources is visible. The grey swath marks the correlation reported by \citet{oyarzun17} for Ly$\alpha$-emitting galaxies, which shows an excellent agreement with our own measurements. Although their stellar mass range is $\log [M_{\rm star}/M_\odot] =7.6-10.6$, the majority lies below $10^{9.5}M_\odot$ and thus should be well matched to the stellar mass range for LAEs \citep{gawiser06,hathi16, santos20, arrabalharo20}. 

In the two right panels, we renormalize the SFR as ${\rm SFR}_{\rm UV,renorm} \equiv C_{\rm UV}\cdot {\rm SFR}_{\rm UV}$. Both panels show identical data which are color-coded by $L_{\rm UV}$ (top) and E$(B-V)$ (bottom), respectively. The renormalization moves our LAEs to the right in better alignment with the HPS sources and the H$\alpha$-Ly$\alpha$ emitters. Only the former are used in the fitting but the latter are rescaled consistently in the figure.

A small fraction (9\%) of galaxies lie significantly outside an otherwise tight sequence formed by the rest, which we demarcate with a dot-dashed line. The outlier group largely consists of high UV-luminosity galaxies ($\gtrsim 10^{29}$~erg~s$^{-1}$~Hz$^{-1}$, or $M_{\rm UV}\lesssim -22.4$) with uncharacteristically blue colors (E$(B-V)\lesssim 0.2$) and near the EW cutoff ($\approx 20$ \AA). Eight of this group belong to the HPS sample,  four are H$\alpha$-Ly$\alpha$ emitters and additional three are in our sample (one in the protocluster region and the other two in the field). Given that both HPS sources and our LAE selection target single line emission, we cannot rule out that some are low-redshift interlopers such as [O~{\sc ii}] emitters at $z=0.34$. In Section~\ref{sec:environment}, we discuss how the exclusion or inclusion of these three LAE candidates affects our conclusion on the environmental dependence on \fesc.

After excluding these outliers, we fit the data to a power-law and find that the scatter is reduced from 0.29~dex, for the unaltered data, to 0.21~dex. When we account for the expected effect of photometric scatter using image simulations, the intrinsic scatter in the \fesc--$C_{\rm UV} {\rm SFR}_{UV}$ scaling relation is less than 0.19~dex.

The physical origin of the luminosity-dependent $C_{{\rm Ly}\alpha}$ is unclear. However, it is worth noting that our result does not mean that $C_{\rm UV}{\rm SFR_{\rm UV}}$ is the true SFR; in such a case, \fesc\ ($\equiv$ SFR$_{{\rm Ly}\alpha}$/SFR$_{\rm UV}$) in the abscissa should also scale accordingly, which would undo the alignment.  

One possibility is that dust laws change with UV luminosity such that more UV-luminous galaxies obey the \citet{cal2000} law while less luminous ones gradually transition to a law more similar to the SMC law. 
\citet{kusakabe15} reported that the average IR luminosity of LAEs at $z=2.2$ measured from the {\it Spitzer}/MIPS and {\it Herschel}/PACS image stacks lie well below that expected under the \citet{cal2000} extinction as evidence that the LAE population may be better described by the SMC law. Similarly, \citet{reddy12} found that  UV-selected star-forming galaxies $z\sim2$ with young stellar population ages ($\lesssim$100~Myr) appear to obey an SMC-like dust law while older galaxies follow the \citet{cal2000} law.

\begin{figure}
\epsscale{1.2}
\plotone{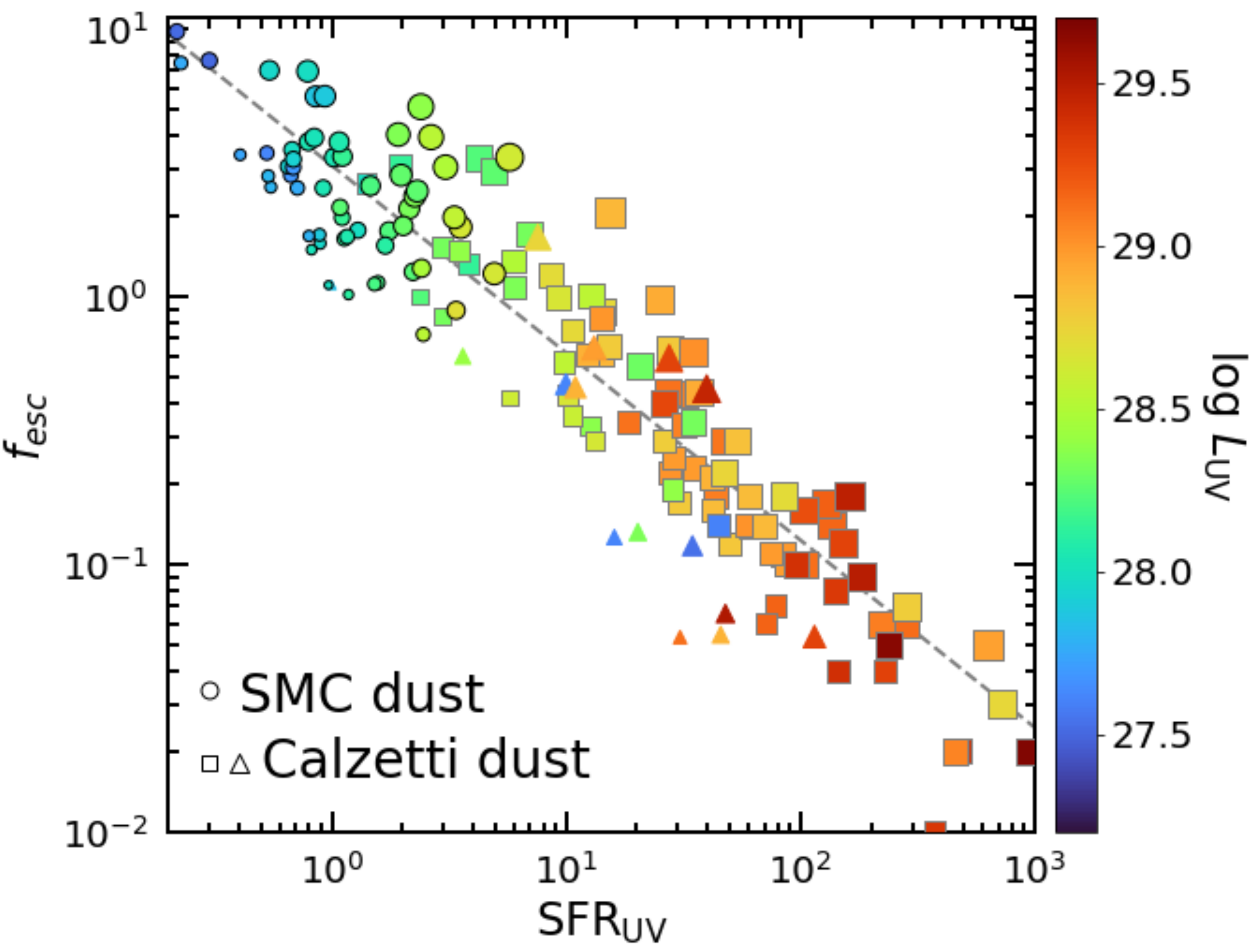}
\caption{ 
Ly$\alpha$ escape fraction as a function of intrinsic SFR$_{\rm UV}$. Both SFR$_{\rm UV}$ and \fesc\ are computed for our LAEs assuming the SMC dust law, while other datapoints are calculated assuming the \cite{cal2000} dust law. Symbols are color-coded with galaxy's observed UV luminosity, with sizes reflecting their Ly$\alpha$ luminosity and shapes identical to those in Figure~\ref{fig:fesc_sfr_corr}.
 }
\label{fig:fesc_sfr_corr2}
\end{figure}

In Figure~\ref{fig:fesc_sfr_corr2}, we show the same data but this time we apply the SMC law for our LAEs while the HPS sources and the H$\alpha$-Ly$\alpha$ emitters still obey the Calzetti law. The tightness of the correlation is comparable to that in Figure~\ref{fig:fesc_sfr_corr} but with fewer outliers. Modeling a luminosity-dependent change in dust laws could conceivably further tighten the scaling relation. The transformation shown in the figure effectively moves our LAEs (in the top left panel of Figure~\ref{fig:fesc_sfr_corr}) to lower SFR (to the left in the x-axis) and to higher \fesc\ values by the same factor. 

One caveat of this interpretation is that it drives already high \fesc\ values even higher. Nearly all of our LAEs would have Ly$\alpha$ escape fractions $\gg$100\%. The problem may be mitigated if low UV-luminosity (or SMC laws) go hand in hand with low metallicity, young ages, or level of burstiness in SFHs: these traits either boost the production of ionizing radiation (and thus Ly$\alpha$ production) or enhances Ly$\alpha$ transmission. A comprehensive study of these properties in LAEs is needed to put to test this hypothesis. \\

\section{Environmental Impacts on the Physical Properties of LAEs}\label{sec:environment}

As detailed in Section~\ref{subsec:protocluster}, our sample includes 24 LAEs residing in a protocluster environment (\citealt{toshikawa16}; \citetalias{shi19b})  offering us a rare opportunity to explore how the environment affects the Ly$\alpha$ properties of the galaxies therein. 
\begin{figure*}
\epsscale{1.1}
\plotone{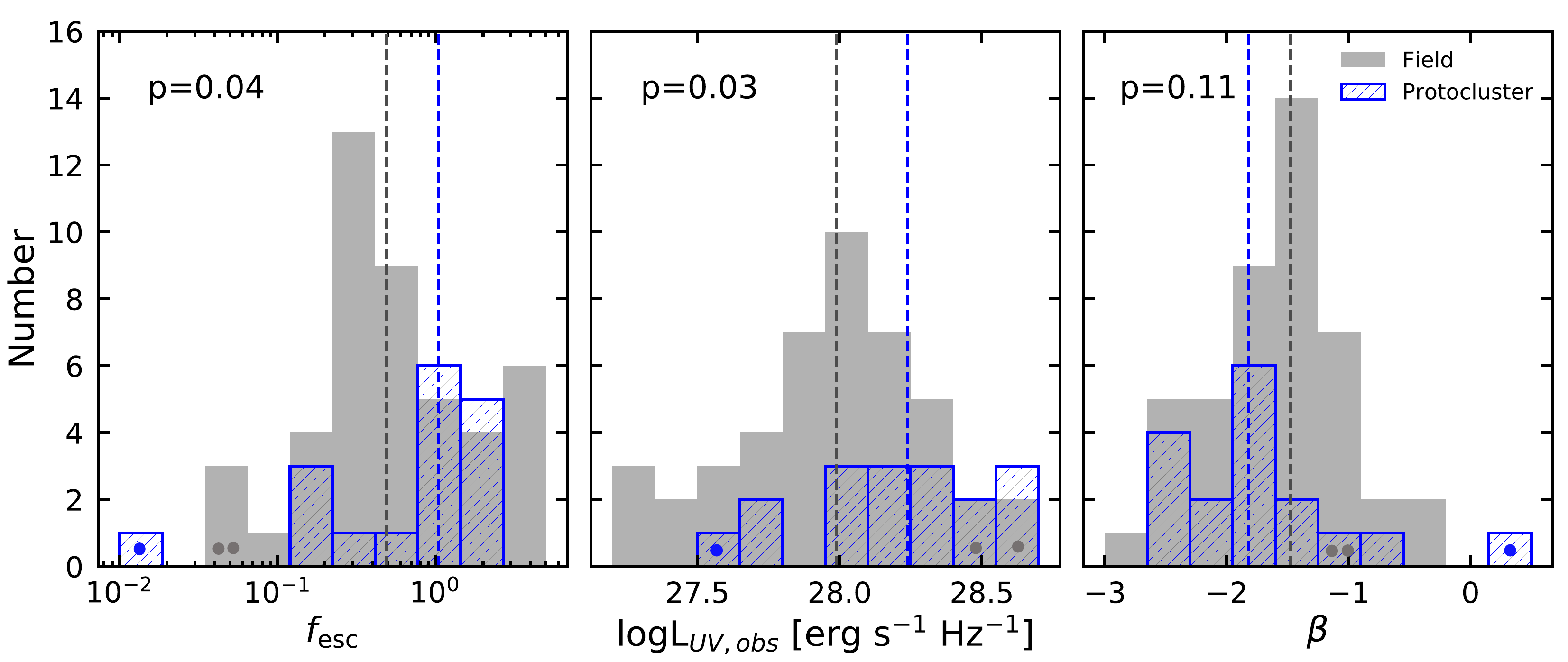}
\caption{ 
Ly$\alpha$ escape fraction, observed UV luminosity, and UV slope $\beta$  distributions of the protocluster LAEs (blue hatched) and the field LAEs (grey filled). Dashed vertical lines measure the median values of each subsample. 
Three sources in our sample that are likely [O~{\sc ii}] emitters (see Section~\ref{subsec:luv_dependece} and Figure~\ref{fig:fesc_sfr_corr}) are marked as filled circles in the relevant bins. 
The $p$ values from the K-S tests noted in each panel include these sources. Excluding them makes it more statistically significant that the protocluster- and field LAE distributions are not drawn from the same parent sample.
}
\label{hist_esc}
\end{figure*}
In Figure~\ref{hist_esc}, we show the \fesc\ distribution split into the `field' (grey) and `protocluster' (blue hatched) groups. The median values are indicated as dashed lines. 
The protocluster members not only show higher \fesc\ values than the field LAEs by a factor of $\approx$2 -- $\langle f_{\rm esc} \rangle=1.05$ vs 0.49 -- but also contain a larger fraction of sources exceeding \fesc$=$1: i.e., 65\% (11/17) vs 33\% (15/45).

The K-S test  results in $p=0.04$ indicating that the two \fesc\ distributions  are distinguishable at a $\approx 2\sigma$ (96\% confidence) level. Despite small-number statistics, our results support the hypothesis that the galaxies' Ly$\alpha$ properties depend on their large-scale environment. 
The middle and right panels of Figure~\ref{hist_esc} show the histograms of $L_{\rm UV}$ and $\beta$ of the same galaxies, the two primary parameters from which \fesc\ is computed. There is a clear tendency towards higher UV luminosities (78\%) and bluer slopes ($\Delta\beta \sim 0.3$) for protocluster galaxies compared to the control group. Similarly, the change in Ly$\alpha$ luminosities roughly tracks that of UV luminosities (see Figure~\ref{escape}). 

A higher UV luminosity in protocluster LAEs (our K-S test yields $p=0.03$) is consistent with that found for another protocluster \citep{dey16}. Here we report, for the first time, a statistical difference in the UV slope $\beta$ for LAEs in different environments. The bluer UV slope, $\Delta \beta \approx 0.3$, corresponds to $\Delta {\rm E}(B-V)\approx 0.07$ assuming the \citet{cal2000} extinction law, leading to a factor of $\approx 1.6$ lower extinction  given everything else equal. This is roughly consistent with the factor of $\approx$2 enhancement we find in \fesc. 
However, the difference in $\beta$ between the two samples is statistically less significant ($p=0.11$) than the other two parameters. It may be owing to the intrinsically narrow range in $\beta$ observed for distant star-forming galaxies\footnote{UV-selected star-forming galaxies with the luminosity range $L_{\rm UV}=(0.1-3.0)L^*$ at $z=3-4$ typically span $\beta$ in [$-2.5,-1.0$] \citep{bouwens09,bouwens12}}. Additionally, photometric estimates of the $\beta$ are expected to have a higher degree of uncertainty because it requires secure detection in two or more bands relative to $L_{\rm UV}$ measurements, which rely primarily on single-band detection.

As discussed in Section~\ref{subsec:luv_dependece}, there are three sources in our LAE sample which are likely to be [O~{\sc ii}] emitters. In Figure~\ref{fig:fesc_sfr_corr}, they lie at the low end of the \fesc\ and $E(B-V)$ values and at the high end of the UV luminosity. One belongs to the protocluster and the other two in the field subsample. These sources are indicated in each panel of Figure~\ref{hist_esc} as filled circles at the bottom of the relevant bin. We repeat the statistical tests after excluding them and found that doing so does not  significantly impact our conclusion but further strengthens it. The $p$ parameter decreases slightly for  \fesc\ values ($p=0.04\rightarrow 0.03$), $L_{\rm UV, obs}$ ($p=0.03 \rightarrow 0.01$), and $\beta$ ($p=0.11 \rightarrow 0.08$). 

To summarize our results, our data suggest strong evidence for higher UV luminosity,  higher line luminosity, and less dust content prevail in galaxies residing in dense protocluster environment. As a result, these galaxies are more efficient producers of Ly$\alpha$ photons and possibly of LyC photons than their field cousins.

It is worth noting that an opposite trend was reported by \citet{lemaux17}, who, based on a spectroscopic sample of star-forming galaxies, concluded that Ly$\alpha$ emission in nine protocluster galaxies in PCl~J1001+0220 -- a protocluster at $z=4.6$ -- is suppressed ($\langle f_{\rm esc} \rangle = 1.8^{+0.3}_{-1.7}$\%) relative to that in the field ($4.0^{+1.0}_{-0.8}$\%). However, the galaxies in the \citet{lemaux20}  sample are significantly more luminous star-formers (with the median brightness $\langle i^\ast \rangle\sim 24.5$) than our LAEs (median $\langle i \rangle \sim 25.8$) and are more massive and dustier than our LAEs. If both results are correct, the implication would be that the environmental trend we observe in our LAEs may be confined to young, low-mass systems and cannot be generalized to more evolved galaxies in the same structure.

\subsection{Physical origins of enhanced \fesc\ in protoclusters}\label{subsec:physical_origin}
Higher Ly$\alpha$ escape fraction for the protocluster LAEs is intriguing. Recent absorption-line studies found that the regions of galaxy overdensities are also H~{\sc i}-richer than average fields \citep{kglee14,kglee18,cai16,liang20}. \citet{liang20} reported a cross-correlation length $4\pm1$~Mpc (comoving) between the LAE overdensity and H~{\sc i} line-of-sight optical depth. These results are broadly consistent with the expectation that both galaxy- and gas overdensities track those of the underlying dark matter. Higher H~{\sc i} column densities at the galaxy scales would result in more frequent random walks of Ly$\alpha$ photons, thereby enhancing the chance of their destruction by dust grains along their paths and thus lowering \fesc\ values. Our result runs counter to this expectation.

One possible explanation is that different ISM conditions for protocluster galaxies facilitate Ly$\alpha$ photons to escape more easily relative to field galaxies. \cite{gazagnes20} studied local star-forming galaxies and found that the escape fraction (for both Ly$\alpha$ and Lyman continuum photons) negatively correlates with  H~{\sc i} covering fraction and dust reddening, favoring the scenario in which low column density channels serve as privileged routes  through which Ly$\alpha$ and LyC photons can escape the galaxy. Such photoionized `tunnels' can be carved out by supernovae feedback (e.g., \citealt{kimm14}) or by the turbulent early phase of the H~{\sc ii} regions and the surrounding molecular clouds \citep{kakiichi19}. In protocluster galaxies embedded in the H~{\sc i}-richer large-scale environment, faster growth brought on by higher star formation efficiency  \citep[e.g.,][]{chiang17} may work through these processes to effectively create a more porous ISM relative to their field counterparts. Such a scenario would have a strong implication for the role of protocluster galaxies in cosmic reionization. 

Another possibility is that we may be witnessing simultaneous births of young primeval galaxies occurring in high-density regions. Extremely young stellar ages may lead not only to a higher level of turbulence in the ISM creating low-density channels \citep{kakiichi19} but also to higher ionizing photon production efficiency (thus increasing Ly$\alpha$ production),  higher specific SFRs \citep{clarke02,endsley20}, and low dust content. Enhanced sSFRs in galaxies residing in high-density environment have been reported recently \citep{shi20,lemaux20}. While the positive SFR-density relation is relatively weak for the general population, there is also tentative evidence that LAEs -- as young, low-mass galaxies -- lie above the  $M_{\rm star}$--SFR main sequence (\citealt{guaita11,hagen16}: but see \citealt{kusakabe18} for a contradictory result) in which such effects may manifest more clearly. 

Finally, protocluster LAEs may be more metal-poor than the field LAEs, producing hotter stellar photospheres and thus higher ionizing radiation efficiency \citep{sternberg03,leitherer10}. While lower metallicity in high-density regions is counter-intuitive, it is plausible if these galaxies are hosted by low-mass ($\lesssim 10^{11}M_\odot$) halos undergoing the first major starburst fueled by the H~{\sc i}-rich environment.  At $z\sim3$, the volume of a single protocluster is immense  \citep{chiang13} and the chemical enrichment therein is expected to be heterogeneous. 

Distinguishing these different but possibly related scenarios would require considerably larger samples of LAEs spanning a range of large-scale environment and deeper imaging data to enable photometric measurements with improved precision.  Optical spectroscopy can enable comparative analyses of the Ly$\alpha$ properties (i.e., the shape and velocity offset) and dust content to test if the distribution of H~{\sc i} gas in the galaxy indeed changes with their large-scale structure. Robust measurements of their SFR, stellar masses, and overall shape of the spectral energy distribution will also help discern a significant difference in metallicity and ages. Deep JWST NIRSPEC observations can place direct constraints on the metallicity of galaxies in a diverse range of environments. 

The One-hundred square-degree DECam Imaging in Narrowbands (ODIN) survey will provide the largest samples of LAEs in protocluster and field environments alike, which can serve as the basis of these investigations. As a new NSF's NOIRLab program (Program ID: 2020B-0201), the survey began in early 2021 to image an area of 91~deg$^2$ with three narrow-band filters sampling redshifted Ly$\alpha$ emission at $z=4.5$, 3.1, and 2.4 (cosmic ages of 1.3, 2.0, and 2.7~Gyr, respectively), straddling the epoch in which the stellar mass assembly rate of both cluster and field galaxies reached its peak \citep{madau14}. Within the survey volume of $\approx$0.24~Gpc$^3$ (comoving), $\approx$100,000 LAEs, $\approx$45 Coma progenitors, and $\approx$600 Virgo progenitor systems are expected. Combined with the existing and upcoming facilities\footnote{The ODIN survey fields largely overlap with legacy fields including the Legacy Survey of Space and Time Deep fields, Euclid deep fields, and one of the  HETDEX survey fields.}, these new datasets will considerably enrich our understanding of the galaxy growth occurring in the largest cosmic structures in direct comparison with that of average galaxies at the same epoch. \\

 \section{Conclusions}\label{sec:con}
 
We have conducted a comprehensive investigation of the Ly$\alpha$ properties of the LAEs at $z\sim 3.1$ including the contribution from the extended, low surface brightness Ly$\alpha$ emission. 
We summarize our results as follows:\\

  -- The average Ly$\alpha$ emission is spatially extended, spanning at least 4\arcsec\ across, in contrast to the UV emission of the same galaxies which is unresolved (Figure~\ref{contours}) confirming  the ubiquity of the Ly$\alpha$ halo  (Section~\ref{sec:lah}). The  Ly$\alpha$ halo has a scale length of $r_{s,h}=4.9 \pm 0.7$ kpc (Figure~\ref{rad77}) and contributes up to $61^{+9}_{-12}$\% of the total line flux in a typical LAE at $z\sim 3$. Protocluster- and field LAEs have similar LAH sizes, implying that large-scale environment is not a major factor that drives the LAH sizes. 
   We provide a simple diagnostic  (Figure~\ref{fraction}) which can help estimate the expected Ly$\alpha$ flux loss  as a function of aperture sizes in a range of seeing values typical in ground-based imaging data. \\

	--	We estimate the Ly$\alpha$ escape fraction for 62 individual LAEs (Section~\ref{subsec:measure}), nearly one-third of which show unphysically high \fesc ($\gtrsim 100$\%). Large \fesc values may be a result of a wide spread in metallicity, age, and star formation histories for the LAEs in our sample; AGN contamination or highly sightline-dependent Ly$\alpha$ emission cannot be ruled out.
	However, the relative  fraction falling into this category is considerably higher for protocluster LAEs than those in the field, hinting at an environmental effect.  After excluding the protocluster LAEs, we find $\langle f_{\rm esc} \rangle\sim$ $40 \pm 26$\% within galaxy-sized apertures. After correcting the flux loss on the average basis, we obtain $\langle f_{\rm esc,tot} \rangle\sim$ $51\pm 31$\%. The modest increase is owing to the compact sizes of the Ly$\alpha$ halo  (Section~\ref{subsec:total_fesc}). Despite the ubiquity of extended line emission in this class of objects, its presence does not warrant a substantial revision to the current picture. \\

	-- Using a compilation of existing measurements, we explore how \fesc\ varies with galaxies' photometric properties (Section~\ref{subsec:var}).  
	Our results suggest that the attenuation of Ly$\alpha$ flux due to interstellar dust is  similar to that experienced by continuum photons ($k_{{\rm Ly}\alpha} \approx k_{1216}$: Section~\ref{sec:dust}), but with a clear difference in their overall transmission by the factor, $C_{{\rm Ly}\alpha}$, which changes with the galaxy's UV luminosity (Figure~\ref{fig:ebv}). $C_{{\rm Ly}\alpha}$ is $\approx$2 for our LAEs, but is reduced to  $\approx$0.5--1 for more UV-luminous galaxies. These results are incompatible with the expectation of several ISM models we have considered, and may support some form of multi-phase interstellar media that allow preferential escape of Ly$\alpha$ photons through selective attenuation (Section~\ref{subsec:ism_models}). \\

	-- We empirically calibrate the Ly$\alpha$ escape as a function of $L_{\rm UV}$ and $\beta$ (Section~\ref{subsec:luv_dependece}, Figures~\ref{fig:fesc_ebv_corr} and \ref{fig:fesc_sfr_corr}). Doing so allows us to predict the \fesc\ value of individual LAEs within $\sim$50\% of the measured value, improving the precision by nearly a factor of 2 compared to the single-parameter ($\beta$) model. The luminosity dependence in the \fesc--$\beta$ relation may result from a gradual shift of the dust law, from SMC-like extinction for low-mass galaxies to the \citet{cal2000} or \citet{reddy16} extinction for more evolved, more luminous galaxies. \\

	-- Protocluster LAEs have higher Ly$\alpha$ escape fractions and are  bluer than their field counterparts (Section~\ref{sec:environment}, Figure~\ref{hist_esc})  suggesting that galaxy formation proceeds differently in dense protocluster environments. We consider different physical scenarios which may explain the observations (Section~\ref{subsec:physical_origin}); these include the possibility that heightened SF activity in the protocluster environment is more conducive to creating a more porous medium facilitating the escape of Ly$\alpha$ and LyC photons. Alternatively, we may be witnessing simultaneous births of extremely young and/or low-metallicity galaxies hosted in low-mass halos in the region. Larger samples of protoclusters and their galaxy constituents combined with sensitive observations from upcoming facilities will place stringent constraints on these scenarios. \\

\acknowledgments
We thank the referee for a careful reading of the manuscript and for suggestions that helped improve this paper.  
We  thank Lucia Guaita and Eric Gawiser for useful comments and suggestions.  YH acknowledges the generous support of the Purdue Research Foundation for completing this work. 
This project is primarily based on observations at Kitt Peak National Observatory at NSF's NOIRLab (NOIRLab Prop. ID 2017B-0087: PI: K.-S. Lee), which is managed by the Association of Universities for Research in Astronomy (AURA) under a cooperative agreement with the National Science Foundation. The authors are honored to be permitted to conduct astronomical research on Iolkam Du'ag (Kitt Peak), a mountain with particular significance to the Tohono O’odham. 
 The work is also based on observations obtained with MegaPrime/MegaCam, a joint project of CFHT and CEA/IRFU, at the Canada-France-Hawaii Telescope (CFHT) which is operated by the National Research Council (NRC) of Canada, the Institut National des Science de l'Univers of the Centre National de la Recherche Scientifique (CNRS) of France, and the University of Hawaii. This work is based in part on data products produced at Terapix available at the Canadian Astronomy Data Centre as part of the Canada-France-Hawaii Telescope Legacy Survey, a collaborative project of NRC and CNRS. This research has been supported by the funding for the ByoPiC project from the European Research Council (ERC) under the European Union's Horizon 2020 research and innovation programme grant agreement ERC-2015-AdG 695561. KS acknowledges the funding from NSFC grant No. 12003023 and the China Postdoctoral Science Foundation grant No. 2020M680086.

\appendix
\section{Robustness of the Dependence of \fesc\ on dust}\label{appendix}

The  escape fraction of Ly$\alpha$ photons is estimated based on three parameters, the reddening parameter (E$(B-V)$) and the line- and continuum luminosity ($L_{\rm UV}$ and $L_{{\rm Ly}\alpha}$) of the galaxy through Equation~\ref{eq6}. Estimation of these parameters comes with varying degrees of uncertainties which can adversely affect our ability to discern the intrinsic scaling relation from our photometric measurements. In this section, we utilize a galaxy simulation to assess the robustness of our photometric measurements and the conclusions based on them presented in Section~\ref{sec:dust}.

We run a simulation containing sources spanning a wide range of relevant parameters. Following the procedure described in \citetalias{shi19b} and \citet{malavasi21}, we use the stellar population synthesis model of   \citet{bc03} to create a model galaxy spectrum (20~Myr in age) characterized by a constant star formation history, the \cite{salpeter} initial mass function, and solar metallicity. Since we are mainly interested in photometric measurements, our results are largely insensitive to these assumptions.
%
To the base spectrum, we assign UV luminosity, UV continuum slope and redshift {\it at random} within the range appropriate for our LAE sample\footnote{$\log L_{\rm{UV}} \in [27, 30]$ erg~s$^{-1}$~Hz$^{-1}$, $\beta \in [-2.1, -0.4]$, and $z \in [3.11, 3.15]$}. The attenuation by the intergalactic media and interstellar dust are corrected using the H~{\sc i} opacity given by \cite{madau95} and \cite{cal2000} dust reddening law, respectively. A Ly$\alpha$ line is then added to the redshifted galaxy SED, which is approximated as a Gaussian profile peaked at $1215.67(1+z)$\AA\ with an intrinsic ${\rm FWHM}=3$\AA\ and a line luminosity in range $\log L_{\rm{Ly}\alpha} \in [42, 44.5]$ erg~s$^{-1}$. We  compute AB magnitudes in $o3$-, $g$-, $r$-, $i$- and $z$-bands by convolving the synthetic SED with the total filter transmission curves. In practice, the line flux falling into a given filter was estimated separately from continuum flux and added to the flux to mitigate the coarse resolution of the \citet{bc03} galaxy templates. Our parent mock catalog consists of a total of 10,000 galaxies whose properties span the full range of the observed values.

From the base catalog, we randomly draw 200 entries and assign position, morphology (disk/bulge and position angle), and size (half-light radii) to each source. Although we use the size distribution consistent with the literature \citep[e.g.,][]{shibuya19},  the angular size of LAEs at $z\sim3$ is small enough  such that most can be considered as point sources in a ground-based image with $\approx$1\arcsec\ seeing as is the case for the present work. These galaxies are added to the science images after being convolved with the image PSF to add realistic photon noise. Source detection and photometric measurements are performed in the identical manner to the real data. We repeat this procedure 100 times, inserting 20,000 artificial sources  to the data.   Of these, 14,273 ($\approx 71$\%) are detected in the $o3$ band at S/N$\geq$7. The  $L_{\rm UV}$, $L_{\rm Ly\alpha}$, $\beta$, and \fesc\ of these $o3$-detected galaxies are estimated following the same procedure as described in \citetalias{shi19b}.

Using the mock photometric catalog, we construct a source list that closely matches the distribution of {\it observed} $L_{\rm UV}$, $L_{\rm Ly\alpha}$, and $\beta$ values of our real LAEs.  Doing so is critical to avoid creating a false sense of agreement or disagreement as any photometric measurement is expected to be more accurate for brighter sources. 
Galaxies are randomly drawn from the catalog with a probability assigned to each source according to its physical parameters. Similar to the real data, only sources with the uncertainty in $\beta$ value less than $0.9$ are retained at this step.  The broad correlation between $L_{\rm UV}$ and $L_{\rm Ly\alpha}$, which stems from the equivalent width cut, is also mimicked by excluding  sources lying outside the region populated by the real LAEs.   Our final mock LAE catalog consists of 1,166 galaxies for which both intrinsic and observed values of the key parameters are available for comparison.

\begin{figure*}
\epsscale{1.1}
\plotone{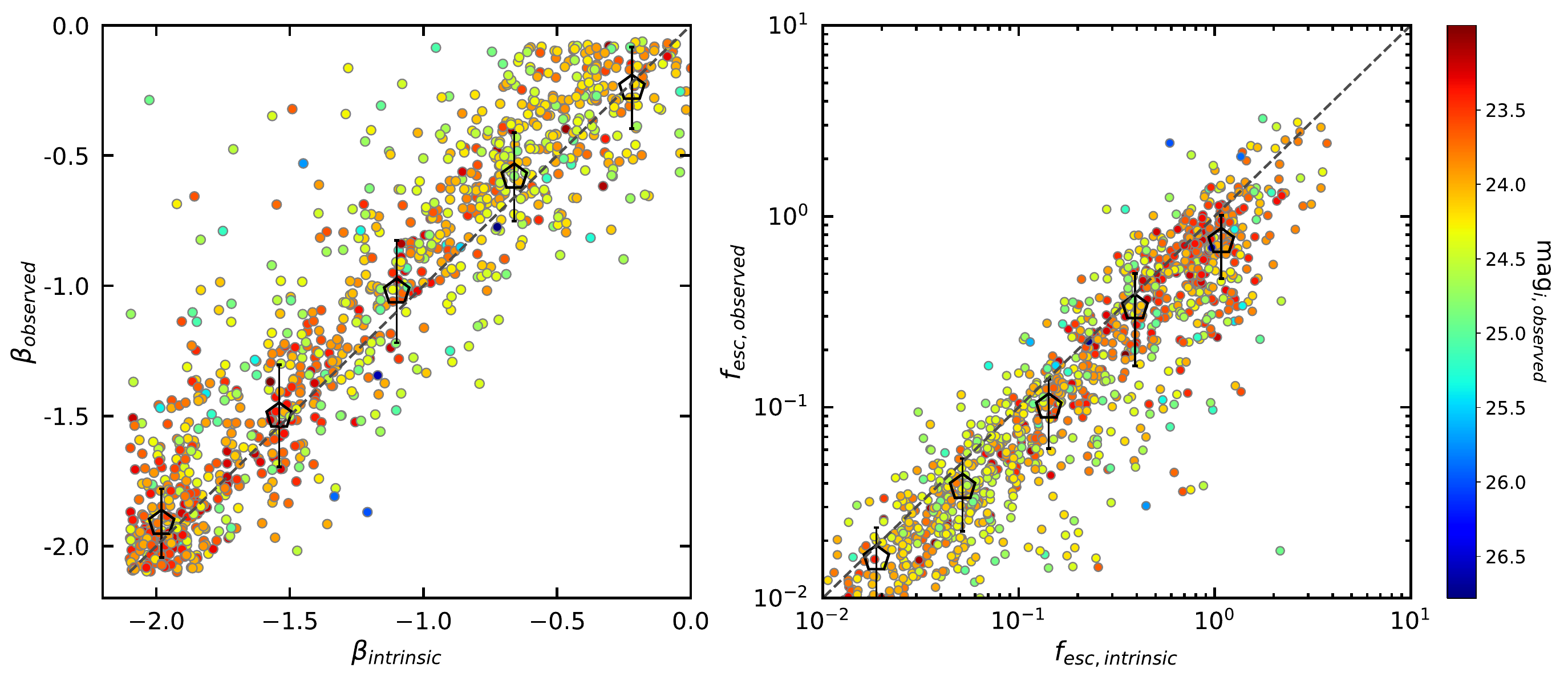}
\caption{ 
The comparisons between the intrinsic and observed values of UV continuum slope $\beta$ and Ly$\alpha$ escape fraction. The synthetic galaxies are color-coded by their observed $i$-band magnitude. In each panel, dashed line marks the one-to-one relationship. Open pentagons represent the mean value in each intrinsic $\beta$/\fesc\ bin after rejecting $>3\sigma$ outliers. 
}
\label{in_out}
\end{figure*}

In Figure~\ref{in_out}, we show the intrinsic and the recovered values of $\beta$ and \fesc. Each galaxy is represented by a circle color-coded by the observed $i$-band magnitude. The overall agreement is evaluated by calculating the mean and the standard deviation of the mock data after rejecting $>3\sigma$ outliers and are shown as open pentagons. Most of these catastrophic outliers tend to be $i$-band faint sources; $i$-bright outliers may simply be the case in which a simulated source falls too close to a real galaxy in the image.
The one-to-one line is shown as a dashed line. The spread in observed $\beta$ values ($\sigma_\beta \approx 0.15-0.20$ in most bins) is primarily due to the photometric scatter in the bands used for the UV slope measurements, which propagates into our \fesc\ determination. The recovered $\beta$ values have a tendency of upscattering to a larger value (i.e., a redder SED), particularly for continuum-fainter sources.  This is likely caused by our $\Delta \beta < 0.9$ cut which, everything else being equal, retains sources which up-scatter to a higher continuum flux while rejecting those which down-scatter. This effect leads to the slight underestimation of \fesc. However, in most bins, the mean is consistent with the intrinsic value within the error. The recovery of the continuum- and line luminosities is much better ($\lesssim 0.1$~dex) than the parameters shown in Figure~\ref{in_out} because the luminosities are  tied to the source brightness in a single band given the narrow redshift range of the LAEs. At the redshift range sampled by the $o3$ filter, the change in luminosity from $z=3.15$ to 3.10 is less than 4\%, much smaller than any measurement uncertainties. All in all, we conclude that our ability to estimate the intrinsic $\beta$ and \fesc\ values is reasonably good. 

\begin{figure*}
\epsscale{1.1}
\plotone{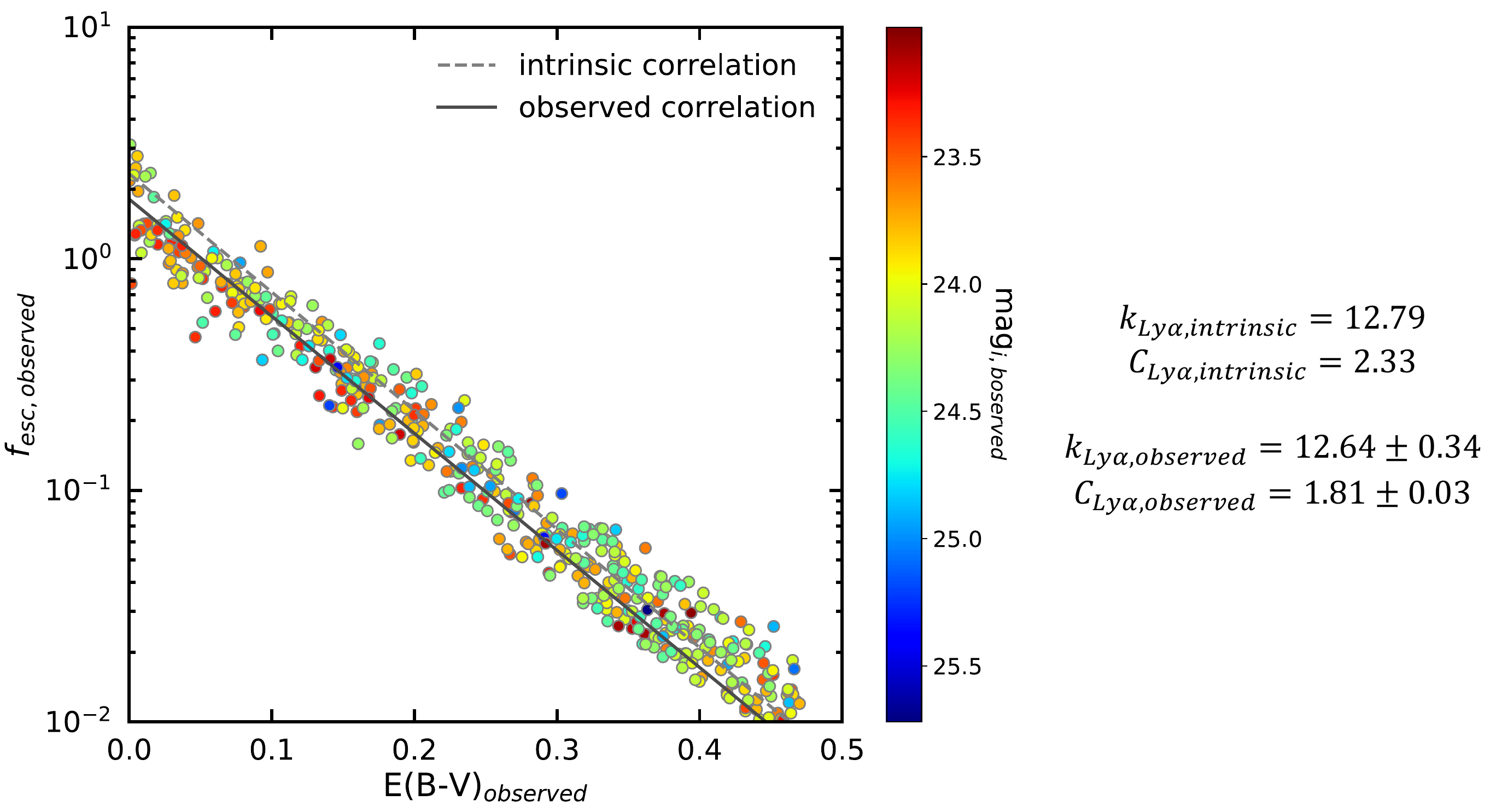}
\caption{  
Using a sample of mock galaxies which obey an intrinsic scaling law between \fesc\ and E$(B-V)$, we test our ability to discern such a relation based on photometric measurements. The scaling law is assumed to be what best describes the D1 LAE sample and is shown as a grey dashed line. Recovered quantities are shown as circles color-coded by the $i$-band brightness for galaxies obeying the scaling law within $\pm 0.15$~dex. The best-fit power-law of these measurements is marked by the solid line. 
}
\label{esc_ebv}
\end{figure*}

We further test our ability to infer the \fesc-E$(B-V)$ scaling relation based on our photometric measurements by defining a subsample of mock LAEs, but this time, only retaining sources whose intrinsic \fesc\ and E$(B-V)$ values obey the dashed line in Figure~\ref{esc_ebv} within $\pm$0.15~dex, which represents the best-fit scaling law (see Table~\ref{tab2} and Section~\ref{sec:dust}). The recovered values are shown as filled circles, once again color-coded by the observed $i$-band brightness. We then determine the best-fit power-law model in the identical manner to the real data and obtain $k_{\rm Ly\alpha}=12.64 \pm 0.34$ and $C_{\rm Ly\alpha}=1.81 \pm 0.03$ indicated by the solid line in the figure. Because we tend to underestimate \fesc\ the most for the bluest (low E$(B-V)$) galaxies, the recovered $C_{{\rm Ly}\alpha}$ is slightly lower than the intrinsic value, while $k_{\rm Ly\alpha}$ is robustly recovered. However, it is clear that the intrinsic and the observed scaling laws are very similar. Thus, we conclude that our measurements can recover the intrinsic scaling law in a statistically robust manner and that the dependence of \fesc\ on interstellar extinction  presented in Section~\ref{sec:dust} is real.

\bibliographystyle{apj}
\bibliography{refs2}  

\end{document}